\documentclass[journal,comsoc]{IEEEtran}

\usepackage{amssymb}
\usepackage{enumitem}
\usepackage{listings, tabu}
\usepackage{color, xcolor}
\usepackage[boxed]{algorithm}
\usepackage{algorithmicx}
\usepackage{url}
\usepackage{colortbl}
\usepackage{tabu}
\usepackage{array, ragged2e}
\usepackage{graphicx,epsfig,epstopdf,amsmath,amssymb,subfigure} 
\usepackage{algpseudocode}

\begin{document}

\title{SARA -- A Semantic Access Point Resource Allocation Service for Heterogenous Wireless Networks}

\author{Qianru~Zhou, 
        Alasdair J. G. Gray, 
        Dimitrios Pezaros, 
        and~Stephen~McLaughlin
\thanks{This research was supported by the EPSRC TOUCAN project (Grant No. EP/L020009/1) and EPSRC NMaaS project (Grant No. EP/N033957/1).}%
\thanks{Qianru Zhou and Dimitrios Pezaros are with the School of Computing Sciences, University of Glasgow, Glasgow, G12 8QQ, U.K. e-mail: (Qianru.Zhou@glasgow.ac.uk; Dimitrios.Pezaros@glasgow.ac.uk).}
\thanks{Alasdair J. G. Gray is with the Department of Computer Science, Heriot-Watt University, Edinburgh, EH14 4AS, U.K. e-mail: (A.J.G.Gray@hw.ac.uk).}
\thanks{Stephen McLaughlin is with the School of Engineering \& Physical Sciences, Heriot-Watt University, Edinburgh, EH14 4AS, U.K. e-mail: (S.McLaughlin@hw.ac.uk).}
}


\maketitle

\pagestyle{empty}
\thispagestyle{empty}

\begin{abstract}
In this paper, we present SARA, a \textbf{S}emantic \textbf{A}ccess point \textbf{R}esource \textbf{A}llocation service for heterogenous wireless networks with various wireless access technologies existing together. By automatically reasoning on the knowledge base of the full system provided by a knowledge based autonomic network management system -- SEANET, SARA selects the access point providing the best quality of service among the different access technologies. Based on an ontology assisted knowledge based system SEANET, SARA can also adapt the access point selection strategy according to customer defined rules automatically. Results of our evaluation based on emulated networks with hybrid access technologies and various scales show that SARA is able to improve the channel condition, in terms of throughput, evidently. Comparisons with current AP selection algorithms demonstrate that SARA outperforms the existing AP selection algorithms. The overhead in terms of time expense is reasonable and is shown to be faster than traditional access point selection approaches.  
\end{abstract}

\begin{IEEEkeywords}

Knowledge based system; access point selection; hybrid wireless network

\end{IEEEkeywords}



\section{Introduction}\label{sec_intro}
Current wireless networks are heterogenous with multiple access technologies coexisting, e.g., WiFi, LTE, 3G/2G, Satellite, Mesh, Ad Hoc, etc. How to make effective use of these technologies, e.g., by choosing which of them will achieve the best connectivity performance based on real time customised requirements, has been the subject of substantial research \cite{nicholson2006improved, vasudevan2005facilitating, fukuda2004decentralized, wu2016two}. However, current access point (AP) selection  technologies usually operate under decentralized control, and are supported by different companies and vendors \cite{nicholson2006improved, vasudevan2005facilitating, fukuda2004decentralized, sato2015resilient, wu2016two, chen2013access}. In most cases, devices produced by one vendor cannot use the network provided by another. There is a lack of a uniformly formatted and centralised knowledge base to enable application of macro-controls over heterogeneous networks within current network management systems \cite{nicholson2006improved}. The development of a high level knowledge base that is universally accepted and machine-processable for network management has been intensively discussed thoughly in the literature \cite{westerinen2001terminology, houidi2016knowledge, zhou2015network}. 

Considerable research has been devoted to the AP selection problem in a homogeneous network (network with one single access technology, e.g., IEEE802.11, LiFi, Satellite). The approach adopted by most of the current network operating systems is to simply choose the AP with the strongest signal strength. However, this signal strength strategy (SSS) does not necessarily guarantee fairness or quality of service to users, particularly in scenarios with an unbalanced load, in which the user and service demands are concentrated in a relatively small area \cite{nicholson2006improved, vasudevan2005facilitating, sato2015resilient, wu2016two, chen2013access}. ``Virgil'', an advanced AP selection strategy was proposed by Nicholson et al. \cite{nicholson2006improved}. It can quickly scan and test all available APs, and select the one with the best connectivity performance. Vasudevan, et al. \cite{vasudevan2005facilitating} considered potential bandwidth as the metric when choosing an access point. 

Although numerous strategies have been proposed for AP selection in homogeneous network, the AP selection in heterogenous networks (where more than one access technologies are adopted) has not yet been as widely investigated. A fuzzy logic approach was applied to select between hybrid LiFi and WiFi networks by Wu et al. \cite{wu2016two}. An extended Analytic Hierarchy Process (AHP) method was adopted in \cite{sato2015resilient} to select and route among the hybrid wireless networks with various technologies (including satellite, 3G, LTE, WiMAX, and Wi-Fi). However, these methodologies only consider fixed metrics, and cannot adapt to complex and dynamic customised rules.

In this paper, we present SARA, a Semantic Autonomic Resource Allocation service. It demonstrates how to choose from different access technologies, such as LTE, LiFi, WiFi, Satellite, and Mesh, considering the real quality of service, and can also adapt to user defined rules by autonomic rules reasoning. It is an application of a knowledge based autonomic network management system -- SEANET (\emph{Semantic Enabled Autonomic management of software defined NETworks}, one of our ongoing work), which use a knowledge base formatted with ontology (a Semantic Web technology, to define the standard vocabulary for a knowledge base) for networks with heterogenous technologies, and provide the capability of autonomic reasoning over the knowledge base. As an application of SEANET, SARA adopts a subset of the ontology used in SEANET, and thus a subset of the knowledge base built for SEANET. The work is evaluated on emulated hybrid wireless networks of different scales.

\section{Knowledge-Based Access Point Selection}
\subsection{Access Point Selection Strategies}
Access point selection strategies are used to choose the access point that will achieve the maximum benefit in channel condition, in terms of throughput, packet loss, etc. 

Current access point selection strategies for one access technology (mainly WiFi) can be summarised as follows:

\begin{itemize}[leftmargin=0em]
\setlength{\itemindent}{0.3in}
\item{\textbf{\emph{Random:}}}  algorithm chooses an AP at random;
\item{\textbf{\emph{SSS:}}} chooses the AP with the strongest signal strength;
\item{\textbf{\emph{Omniscient:}}} simulates an algorithm which uses the results of AP probes to choose the AP with the best bandwidth \cite{nicholson2006improved}.
\end{itemize}

The methodologies proposed for AP selection in homogeneous networks are numerous as the sand on the beach. However, the selection strategy among networks with hybrid technologies are relatively nascent. By improving the Analytic Hierarchy Process (AHP) decision making algorithm, Sato et al \cite{sato2015resilient} designed a disaster resilient network based on Software Defined Networking (SDNing) technology, to automatically select the best network among different types of wireless access technologies. By executing the advanced AHP methodology repeatedly to evaluate the network status, the system can automatically perform handover to the other access networks which have enough available network resources, even if some network node and line failures occur in a part of the network infrastructure, and thus provide persistent communication capability \cite{sato2015resilient}. Wu et al \cite{wu2016two} use fuzzy logic to design an two-stage AP selection method for hybrid visible light communication (VLC)/radio-frequency (RF) networks. The proposed method first determines the users that should be connected to the RF system, and then assigns the remaining users as if in a stand-alone VLC network.

\begin{figure}[hbt]
\centering
\includegraphics[width=0.43\textwidth]{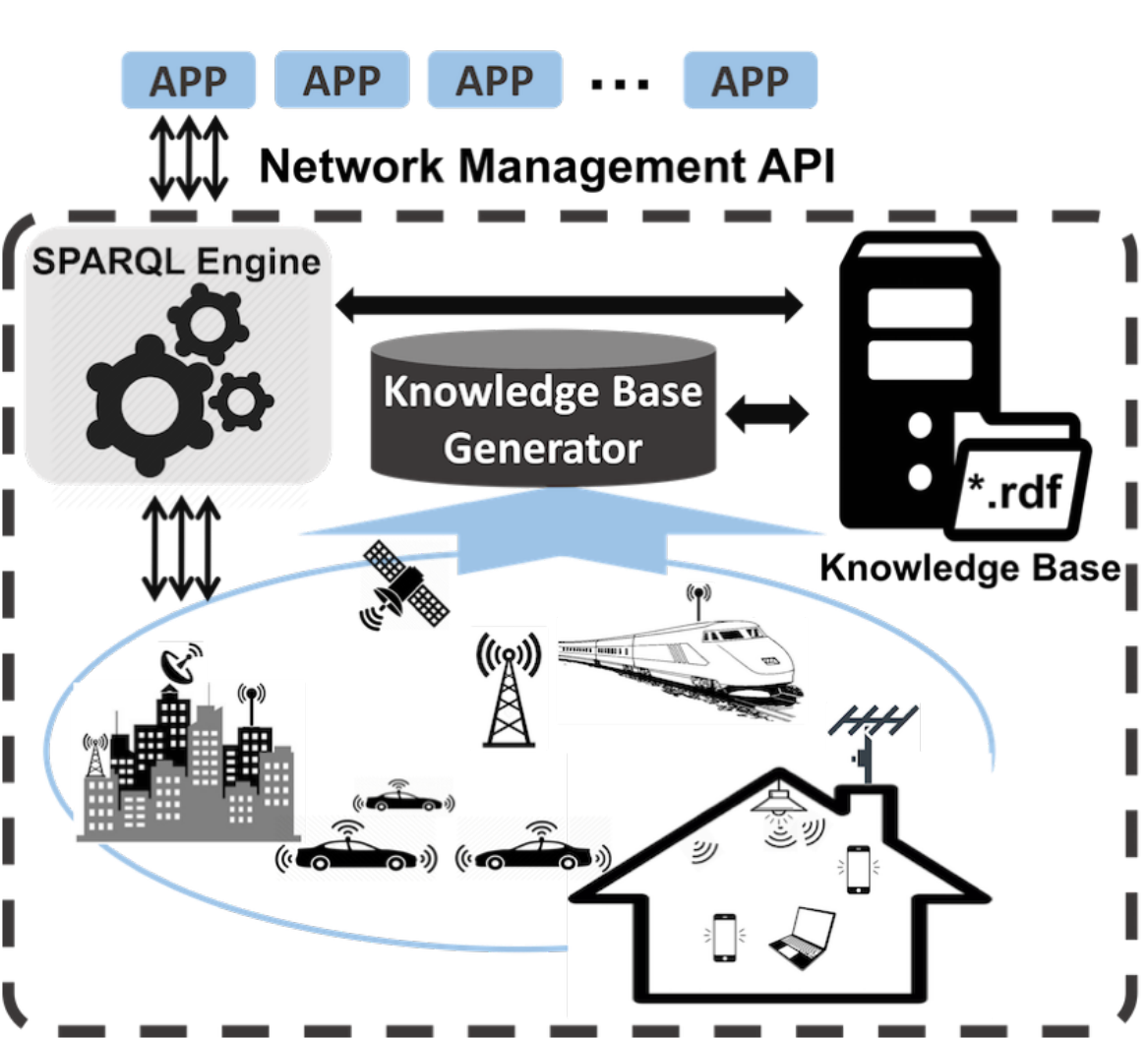}
\caption{Architecture of the SEANET, with the proposed components: knowledge base generator, SPARQL engine and a network management API over the heterogenous networks.}
\label{fig_seanet} 
\end{figure}

\subsection{Knowledge-Based Network Management System}
The SARA application depends entirely on a knowledge based autonomic network management system. The knowledge-based network management system is a software solution that adopts and infers from a knowledge base to solve problems automatically in the network management domain \cite{180}. There are three components in a knowledge-based system: a \emph{knowledge base}, an \emph{inference engine}, and a \emph{user interface}. Central to any knowledge-based system is the knowledge base, which contains all the domain's knowledge as a collection of facts, or in other words, all that is known in a domain. A knowledge-based system requires formally structured data for its knowledge base, not just the traditional databases with only numeric and literal records, but also pointers direct to other objects which in turn have more pointers. In other word, these data are linked \cite{180}. The ideal representation for linked data is an \emph{ontology}. Our prototype knowledge based system SEANET uses an ontology (the standard format for knowledge description in Semantic Web technologies) to build a linked knowledge base (or knowledge graph, which is more commonly used in the community) of the network, which enable self-reasoning. An \emph{inference engine} is able to manipulate the existing knowledge base according to customised rules and come to the conclusions \cite{180}.

Among the various reasons for adopting knowledge based system, the most vital reason is the flexibility. With the assistance of the ontology, the system can achieve competency swiftly by being told or learning new knowledge of a particular domain, and are able to accept new tasks in the form of customised rules, besides, they can adapt to the environment timely by updating the knowledge base in real time \cite{180}. 

As an autonomic network management system, SEANET is designed to perform network management tasks without intervention from external software or a human. It should also be able to close the control loop and adapt to the changing environment by itself [7].

The overall architecture of SEANET is shown in Fig. \ref{fig_seanet}. It consists of three components: a \emph{knowledge base generator}, a \emph{SPARQL (an SQL-like query language which can query over linked-data) engine}, and a \emph{Network Management API}. The \emph{knowledge base generator} is designed to bring knowledge base harmonisation into reality by retrieving unstructured data from nodes in the network, and translating and formatting them into the knowledge base with the ToCo ontology\footnote{\url{http://purl.org/toco/}}. The generated \emph{knowledge base} is technology independent, which can work on networks with hybrid technologies and various topologies and scales. The function of the \emph{SPARQL engine} is to run the query on the generated knowledge base, which is also network technology-independent. Its input is a query string, and the output is a formatted result set. To make the functions of SEANET available for users without knowledge of either telecommunications or Semantic Web technologies, an \emph{API} is provided offering a range of functions from knowledge base generation, query execution, to the execution of network management tasks. With the linked knowledge base and inference rules, SEANET can obtain the abstract knowledge of the network and subsequently issue tasks at an abstract level, hiding the detailed technology--specific operations to itself.  

The generated \emph{knowledge base} can be manipulated using the \emph{SPARQL engine} and thus can be used to answer high level questions such as, \emph{``Which access point is the least busy one (with the lowest number of connected clients)?''}, \emph{``Find me the hosts in the network which are set to drop their flows.''} or \emph{``How many access points are in the neighbourhood of zone Z, and find me the one with the best channel situation defined by the customised rule $R_1$.''} The following section gives an detailed demonstration of adopting the knowledge base and inference rule to implement SARA -- an access point re-selection algorithm for heterogeneous networks.

\subsection{SARA -- Semantic Access Point Resource Allocation Service}
\begin{figure}[hbt]
\centering
\includegraphics[width=0.46\textwidth]{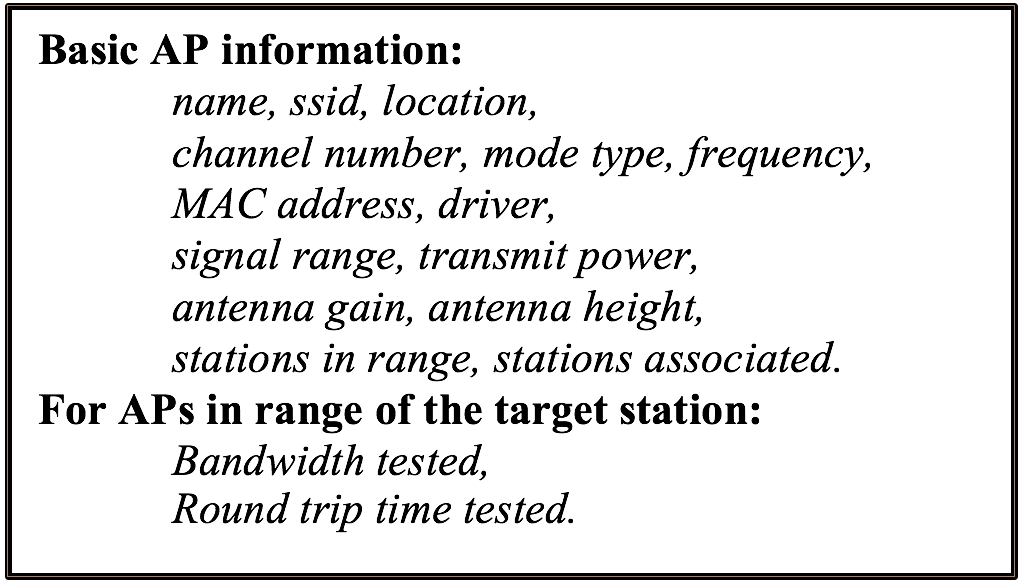}
 \caption{Knowledge entities about Access Point in the SEANET knowledge base.}
 \label{fig_entities} 
\end{figure}

\begin{figure}[hbt]
\centering
\includegraphics[width=0.47\textwidth]{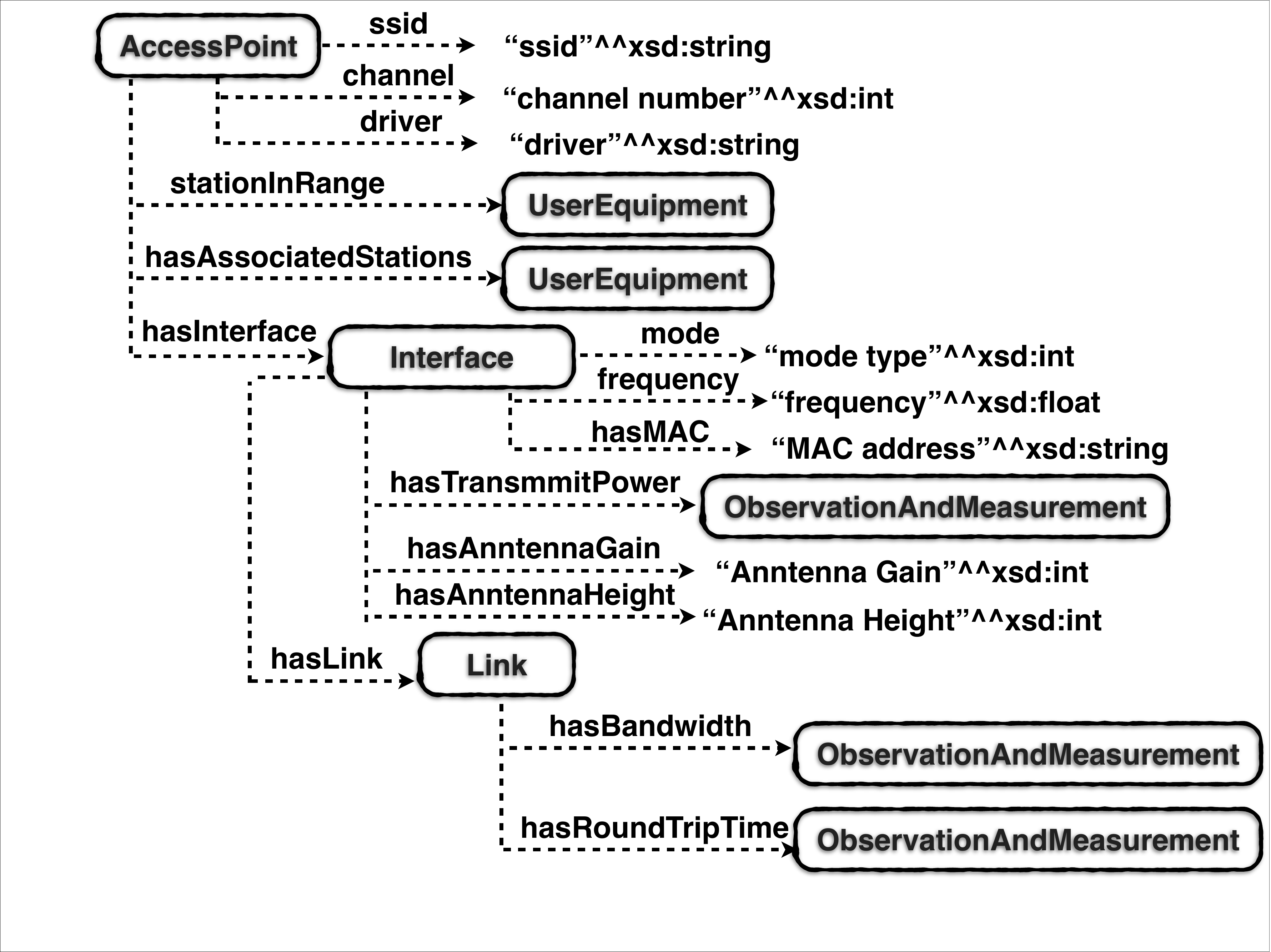}
 \caption{The part of the ToCo Ontology used in SARA.}
 \label{fig_onto} 
\end{figure}

The main knowledge entities used in SARA are shown in Fig. \ref{fig_entities}. As SARA is designed as an application of an ontology-assisted knowledge based system SEANET, the ontology adopted in SEANET, ToCo Ontology, is also used in SARA, as shown in Fig. \ref{fig_onto}.

The goal of SARA is to reallocate the data traffic burden to less-busy APs. It is worth mentioning that the criteria and algorithms for achieving that goal might vary with different scenarios and networks. What SARA tries to demon in this paper is the ability to execute high-level algorithm and customised rule, with an ontology-driven knowledge base. After consulting previous related work on heterogeneous network AP selection, we consider the following criteria in our AP selection strategy, but these criteria and the algorithm follows are not the fixed only solution.

\begin{itemize}[leftmargin=0em]
\setlength{\itemindent}{0.3in}
\item{\textbf{\emph{Number of Current Users:}}} the number of mobile stations associated with an AP. In our post-incident scenario, the more users associated with an AP denotes the heavier traffic load it bears;
\item{\textbf{\emph{Bandwidth:}}} the bandwidth between the AP and the target station, tested with \emph{iperf} automatically by SARA;
\item{\textbf{\emph{Transmit Strength:}}} the signal strength the user received from the AP. It is presented as antenna gain and transmit power in the current SEANET knowledge base.
\end{itemize}

\textbf{SARA algorithm}: SARA's algorithm for selecting a new access point is as follows:

\begin{enumerate}[label=\arabic*]

\item Select all available access points in range.

\item Test the bandwidth between the target station and the available access points. Update this information in the SEANET knowledge base.

\item Query the knowledge base for the following properties of these APs.
	\begin{itemize}[leftmargin=0em]
	\setlength{\itemindent}{0.3in}
		\item[-] The total number of mobile stations currently associated with the AP;
		\item[-] Bandwidth between the station and these APs;
		\item[-] The transmit power of the AP;
		\item[-] The antenna gains of the AP.	
	\end{itemize}

\item Sort the APs by these three properties with following priority: total number of stations $>$ bandwidth $>$ transmit power $>$ antenna gain.

\item Choose the best AP, i.e. the first in the sorted list.
\end{enumerate}

This algorithm is implemented as a SPARQL query on the knowledge base SEANET build. The SPARQL query is shown in Algorithm \ref{listing_sparql}.

\begin{algorithm}
\caption{SPARQL query to find the best AP for a mobile station ``sta1'' considering the number of traffic load and signal strength of the APs.}
\label{listing_sparql}
\begin{algorithmic}
\State{PREFIX net:~\url{<http://purl.org/toco/>}}
\State{SELECT  ?aps (COUNT(?asso) AS ?cnt)   } 
\State{WHERE \{ }
\State{\quad    ?aps :stationsInRange :sta1;}
\State{\quad\quad\quad    :associatedStations ?asso; :hasWLAN ?w. }
\State{\quad    ?w :antennaGain ?g; :hasTxPower ?tx. } 
\State{\quad    ?asso a :Association; :From ?aps; :To :sta1; }
\State{\quad\quad\quad    :hasBandWidth ?bw.}
\State{\quad    ?bw :hasValue ?bwValue.}
\State{\quad     \}}
\State{\quad    GROUP BY ?aps}
\State{\quad    ORDER BY ?cnt DESC(?bwValue) DESC(?g) DESC(?tx)}
\end{algorithmic}
\end{algorithm}

\textbf{Service specific AP selection rule}: In addition to the AP selection algorithm users can set up their customised rules to select specific technology for different kinds of service. A example of the rules in first order logic is shown below.

$$\forall UserEquipment(?u) \land hasService(?u, ?s) $$
$$ \land  isVideo(?s) \to associateTo(?u, LTE)$$

The customised rule can be reasoned automatically by SEANET. If the rule is enabled, the AP will be assigned according to the rule, the rest are assigned by the default AP selection algorithm. The evaluation of the service specific AP selection rule will be presented as a separate work, thus is not included in this paper.

\begin{table}[ht]
\tiny
\centering
\caption{Simulation Parameters. }
\label{table_para}
\resizebox{0.46\textwidth}{19mm}{
\begin{tabular}{|c|c|}
\hline
\textbf{Variable} & \textbf{Value} \\ \hline
WiFi (IEEE 802.11b) AP range & 75m \\
Satellite AP range & 1000m \\
LTE AP (base station tower) range & 1000m \\
Mobile Station range & 75m \\
WiFi (IEEE 802.11b) Wireless Channel Bandwidth & 20 Mbits/s \\
Satellite Channel Bandwidth & 10 Mbits/s \\
LTE Channel Bandwidth & 10 Mbits/s \\
Default Packet Loss & 10\% \\
Mean Mobile Stations speed & $2.5$ m/second \\
Simulation area & $300$m $\times$ $300$m \\ 
Simulation time & $1000$s \\
\hline
\end{tabular}
}
\end{table}

\begin{figure*}[htbp]
\centering
\subfigure[]{ \label{fig_move:subfig:a}  
\includegraphics[width=1.6in]{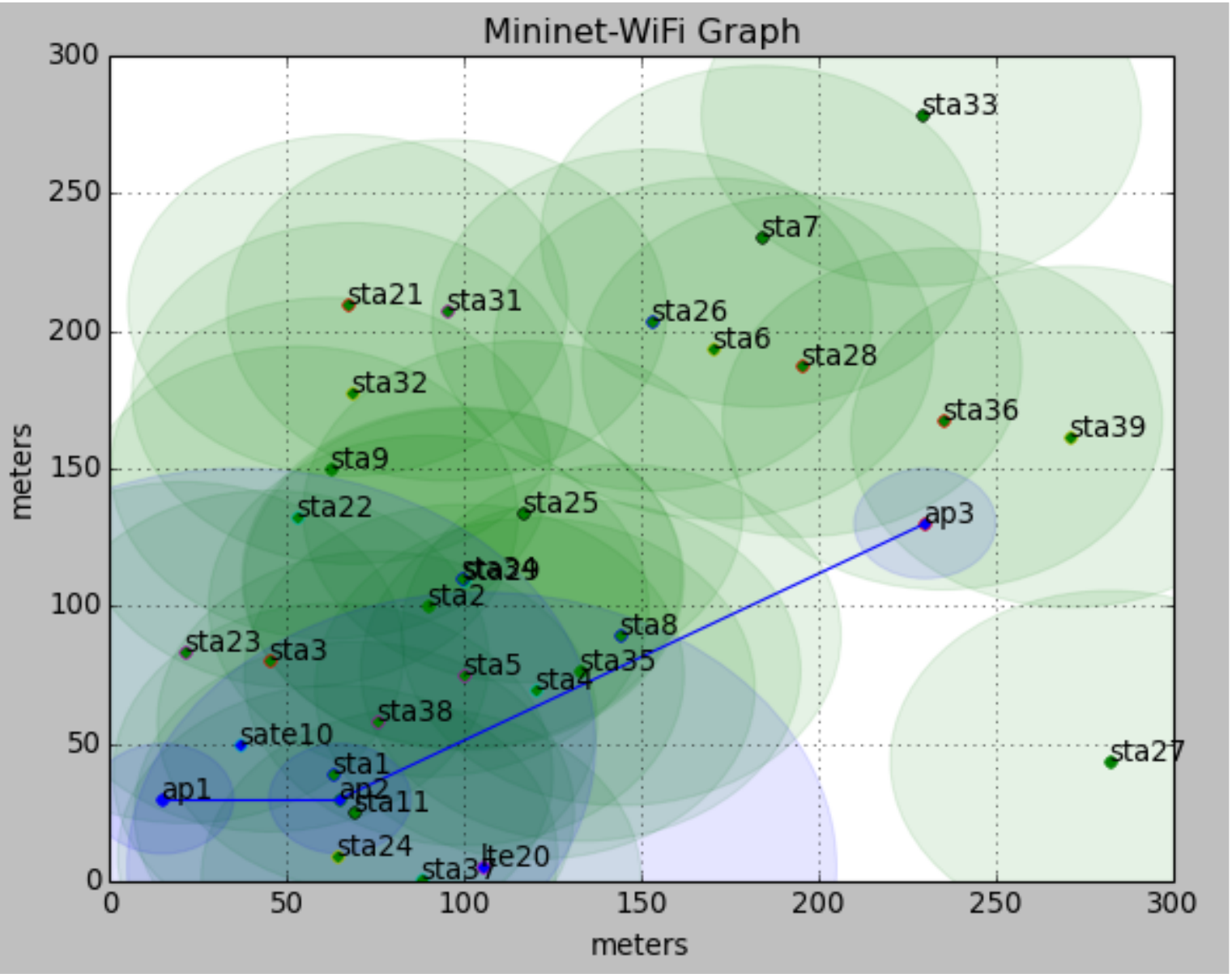}
} 
\subfigure[]{ \label{fig_move:subfig:b}  
\includegraphics[width=1.6in]{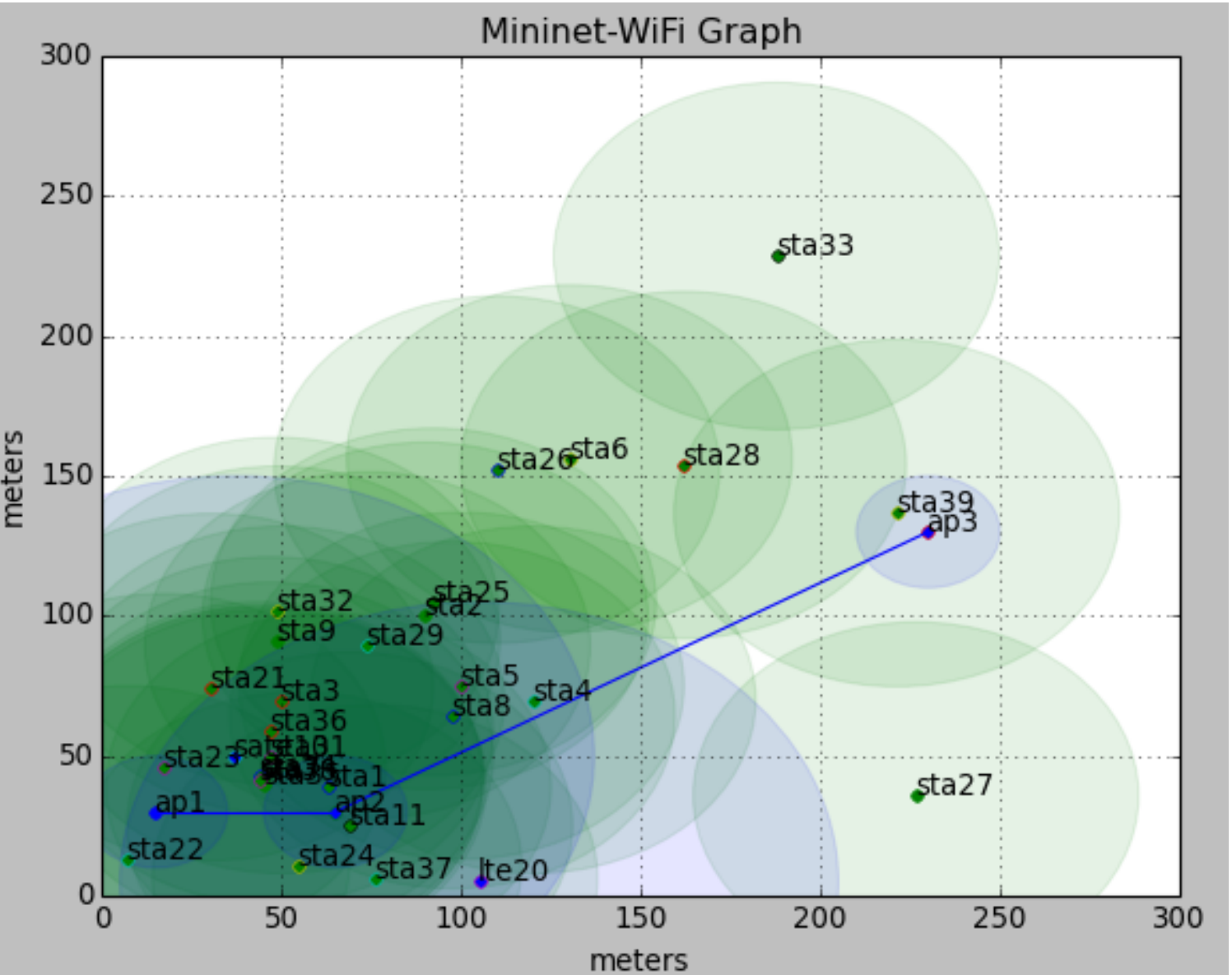}
} 
\subfigure[]{ \label{fig_move:subfig:c}  
\includegraphics[width=1.6in]{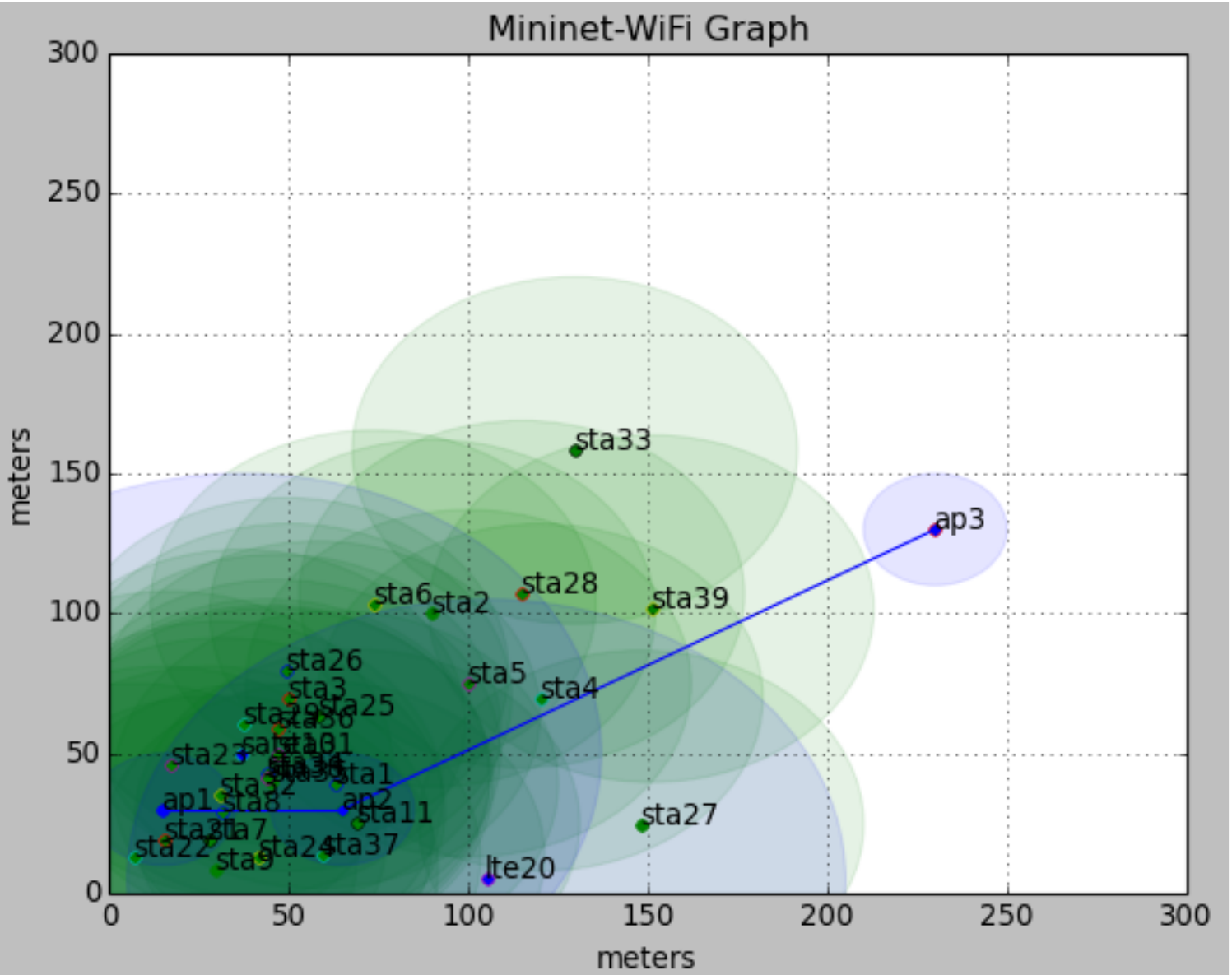}
} 
\subfigure[]{ \label{fig_move:subfig:d}  
\includegraphics[width=1.6in]{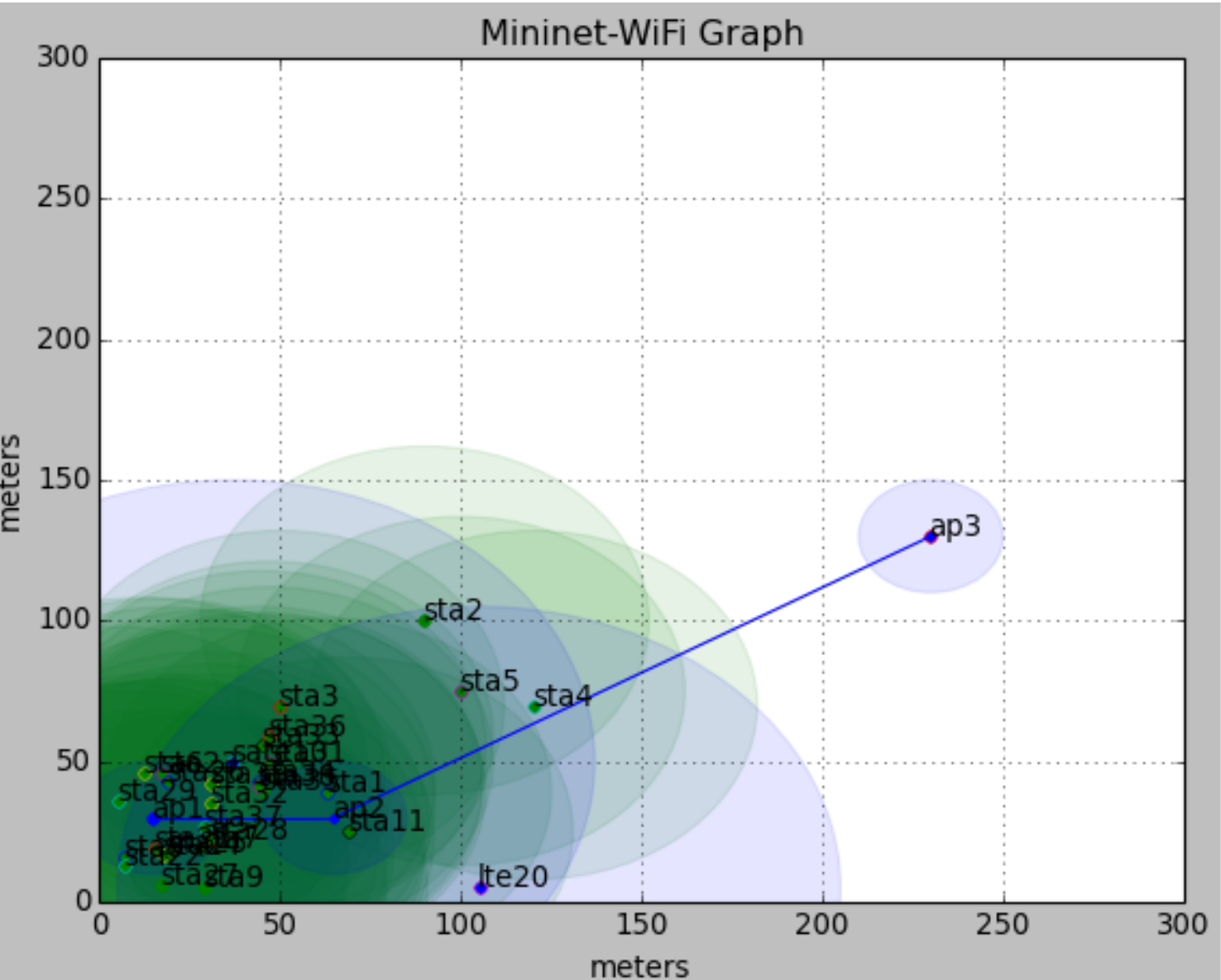}
} 
\caption{The movement of 30 mobile stations at different time point.}
\label{fig_move}
\end{figure*}

\begin{figure*}[htbp]
\centering
\subfigure[Sta 1]{ \label{fig_throughtput:subfig:a}  
\includegraphics[width=2in,height=1.3in]{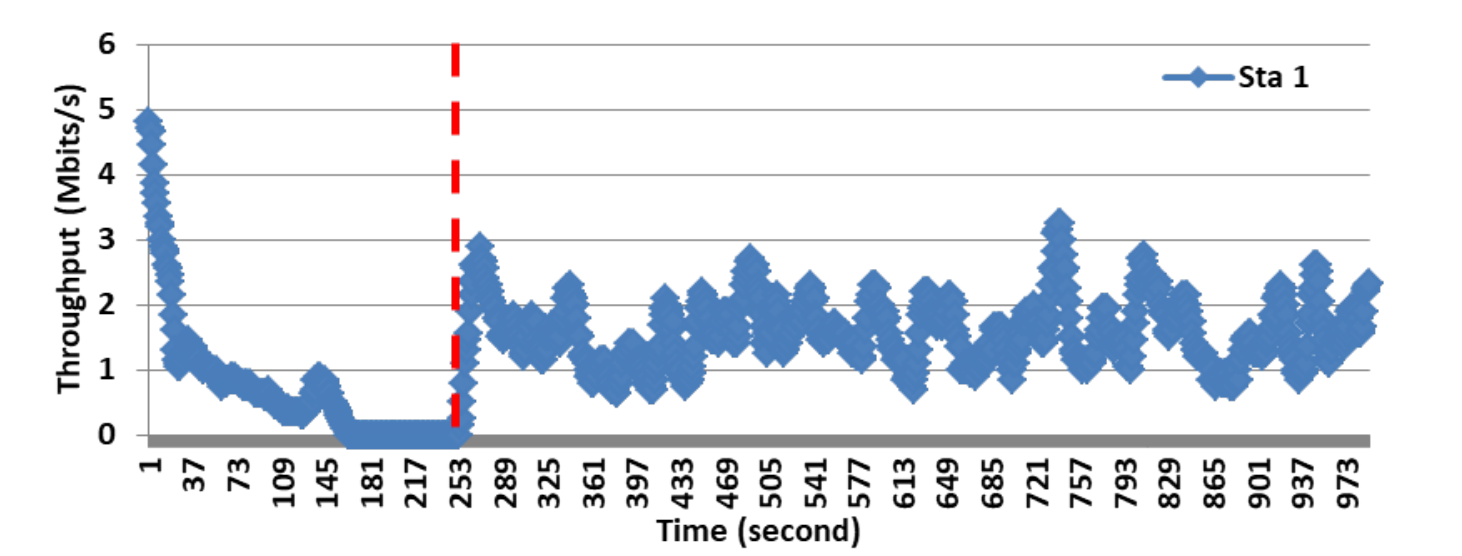}
} 
\subfigure[Sta 2]{ \label{fig_throughtput:subfig:b}  
\includegraphics[width=2in,height=1.3in]{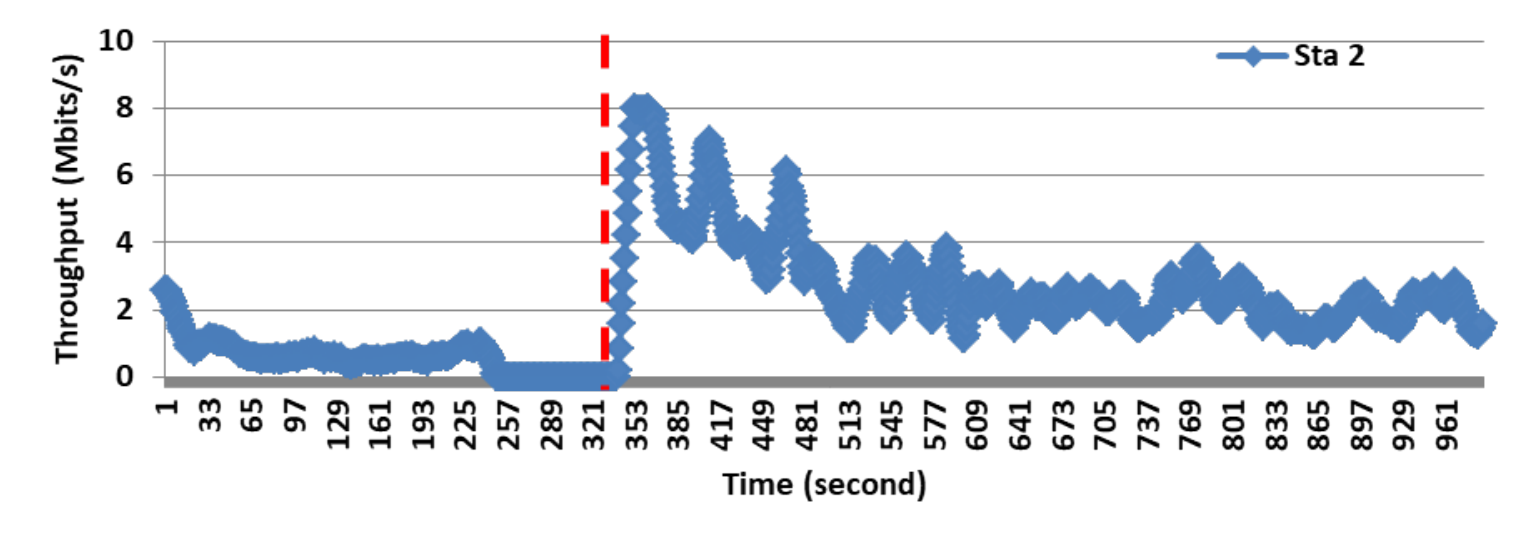}
} 
\subfigure[Sta 3]{ \label{fig_throughtput:subfig:c}  
\includegraphics[width=2in,height=1.3in]{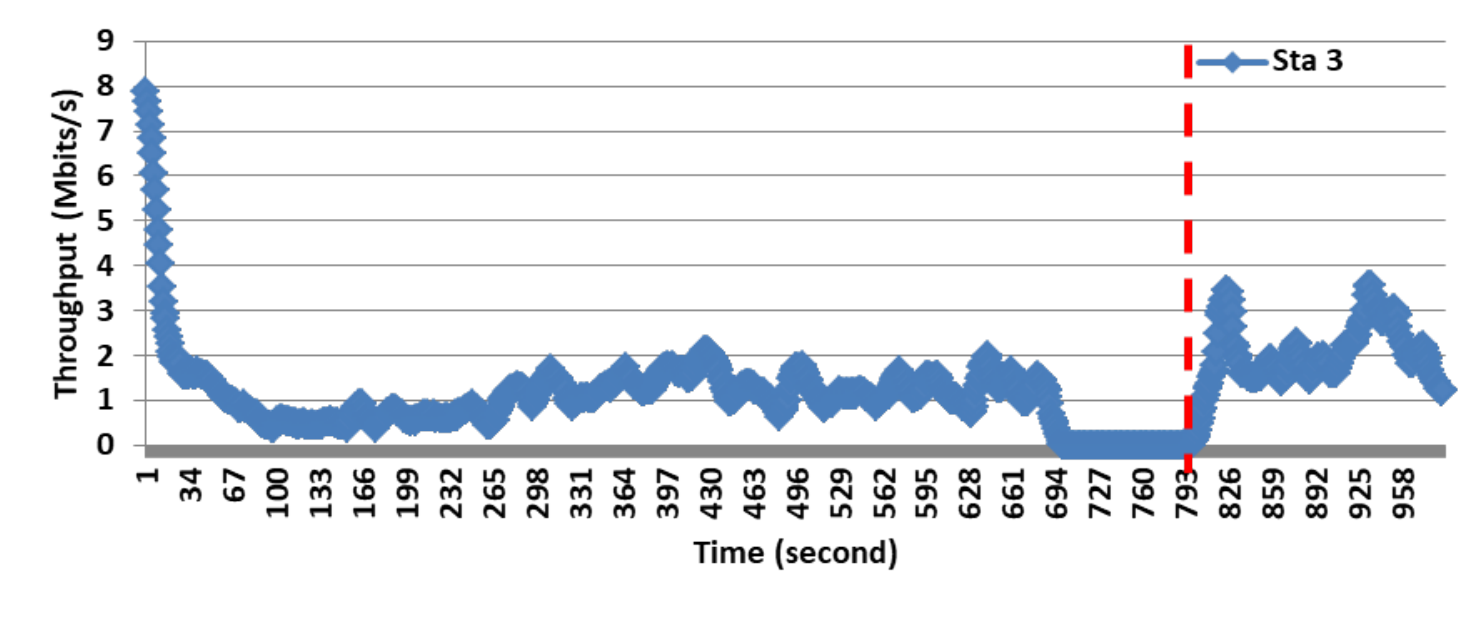}
} 
\subfigure[Sta 4]{ \label{fig_throughtput:subfig:d}  
\includegraphics[width=2in,height=1.3in]{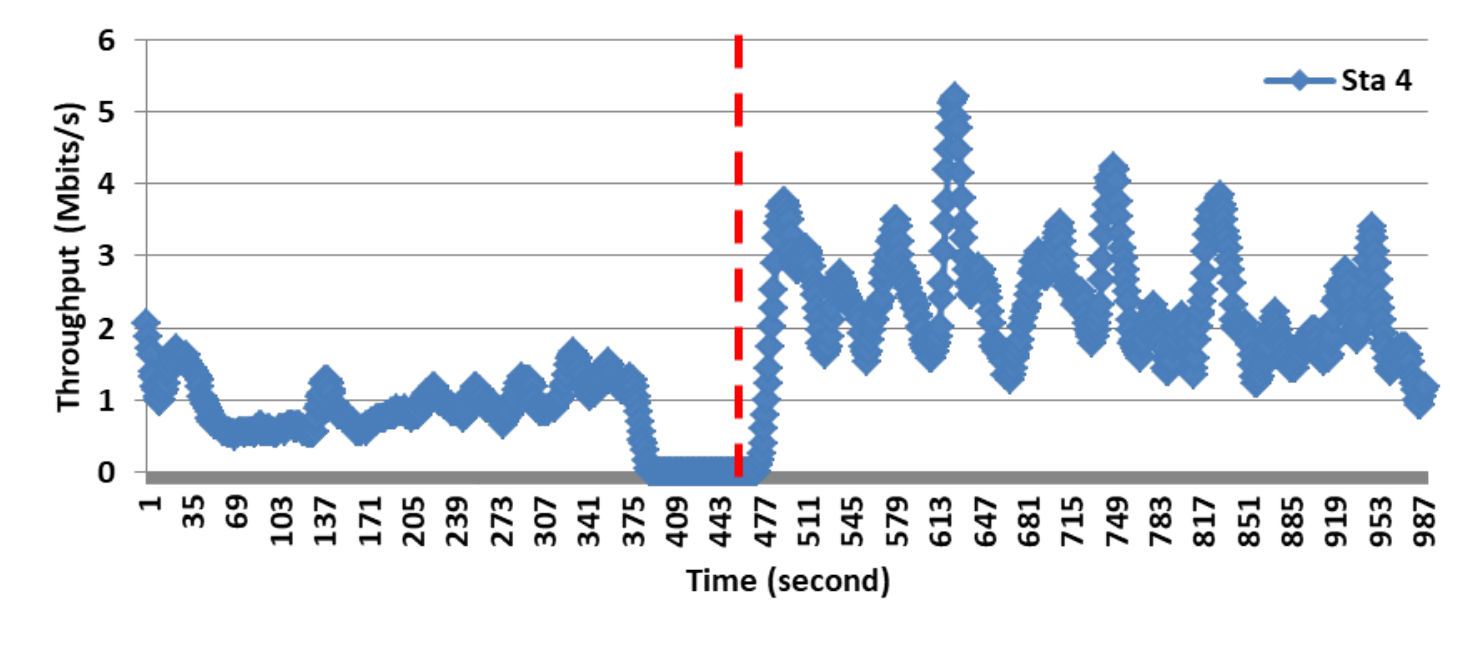}
} 
\subfigure[Sta 6]{ \label{fig_throughtput:subfig:e}  
\includegraphics[width=2in,height=1.3in]{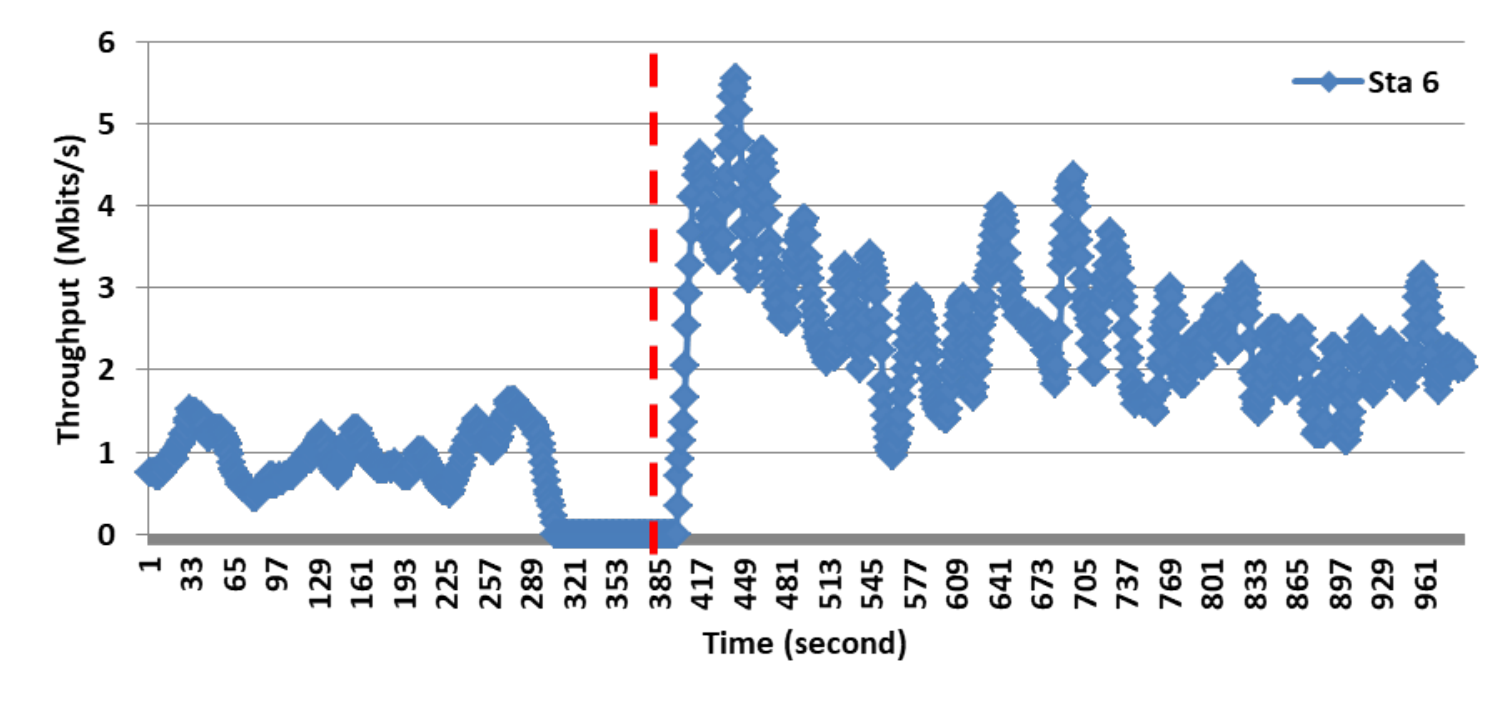}
}
\subfigure[Sta 7]{ \label{fig_throughtput:subfig:f}  
\includegraphics[width=2in,height=1.3in]{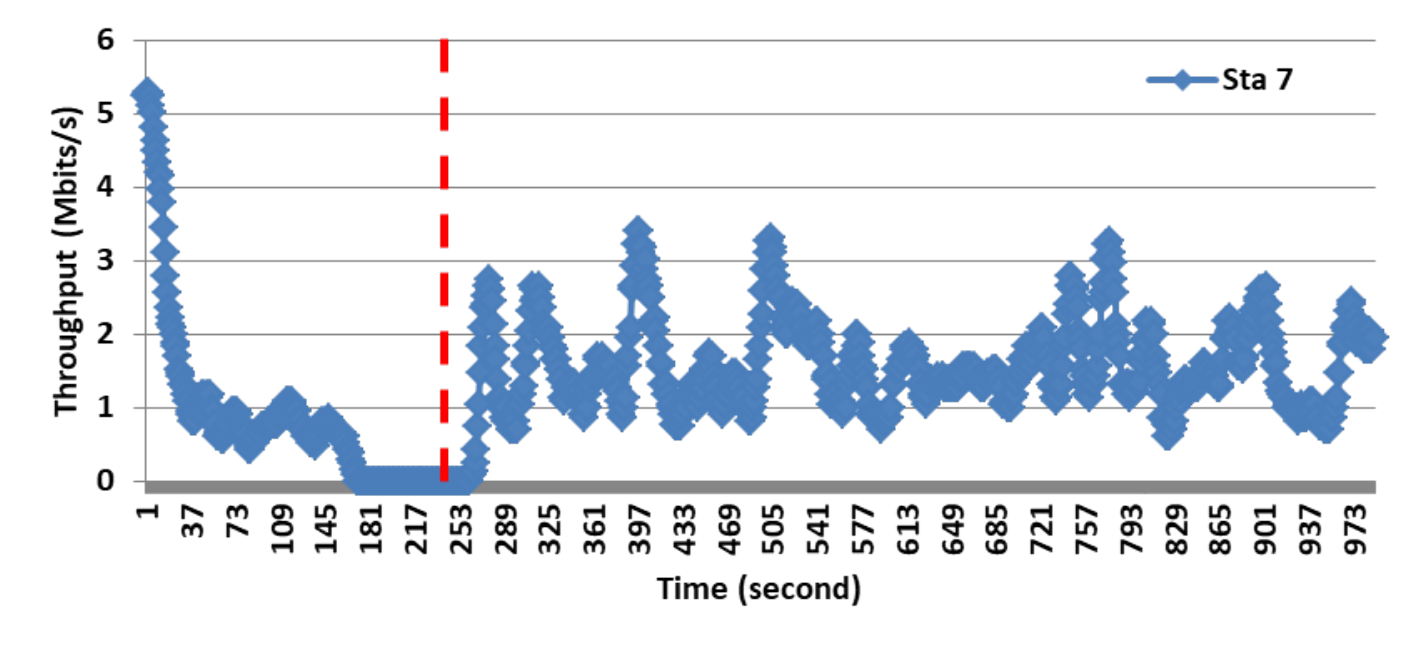}
}
\subfigure[Sta 10]{ \label{fig_throughtput:subfig:g}  
\includegraphics[width=2in,height=1.3in]{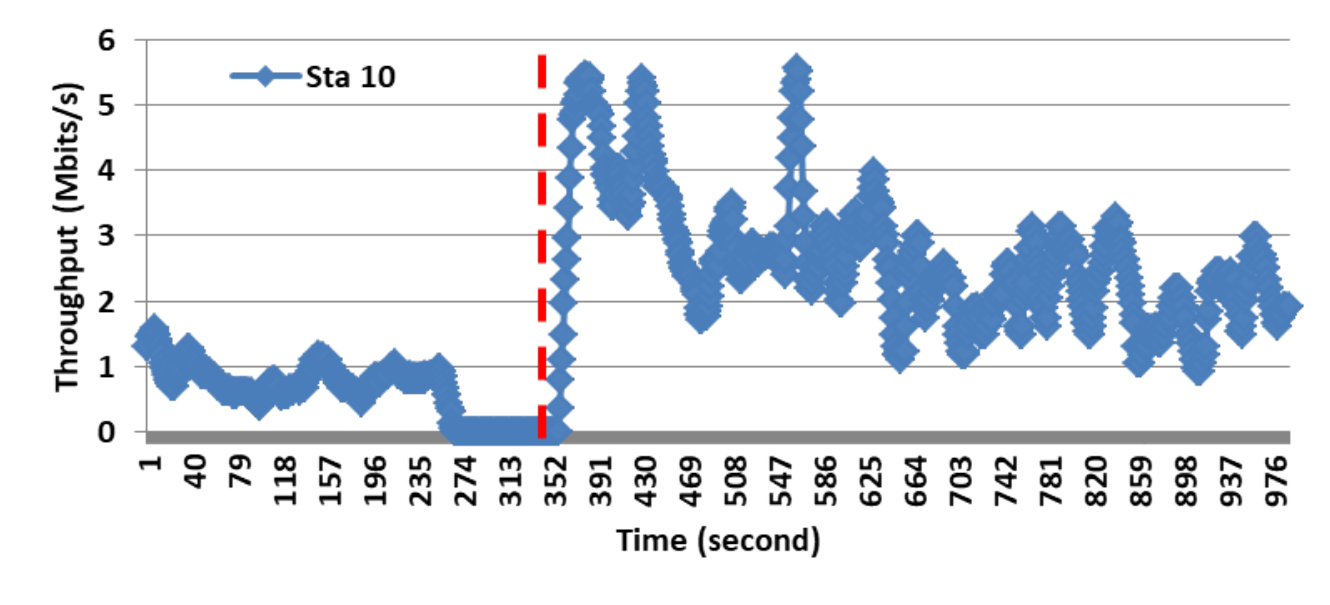}
}
\subfigure[Sta 12]{ \label{fig_throughtput:subfig:h}  
\includegraphics[width=2in,height=1.3in]{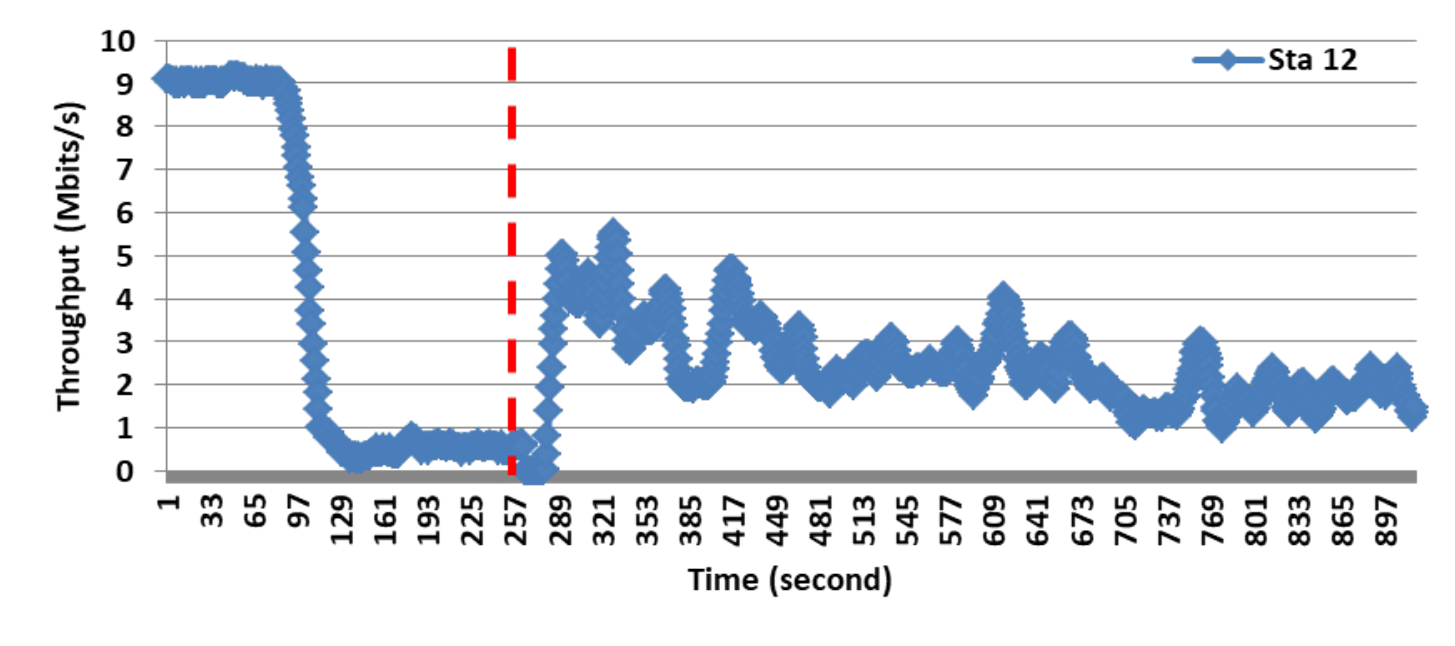}
}
\subfigure[Sta 15]{ \label{fig_throughtput_1:subfig:i}  
\includegraphics[width=2in,height=1.3in]{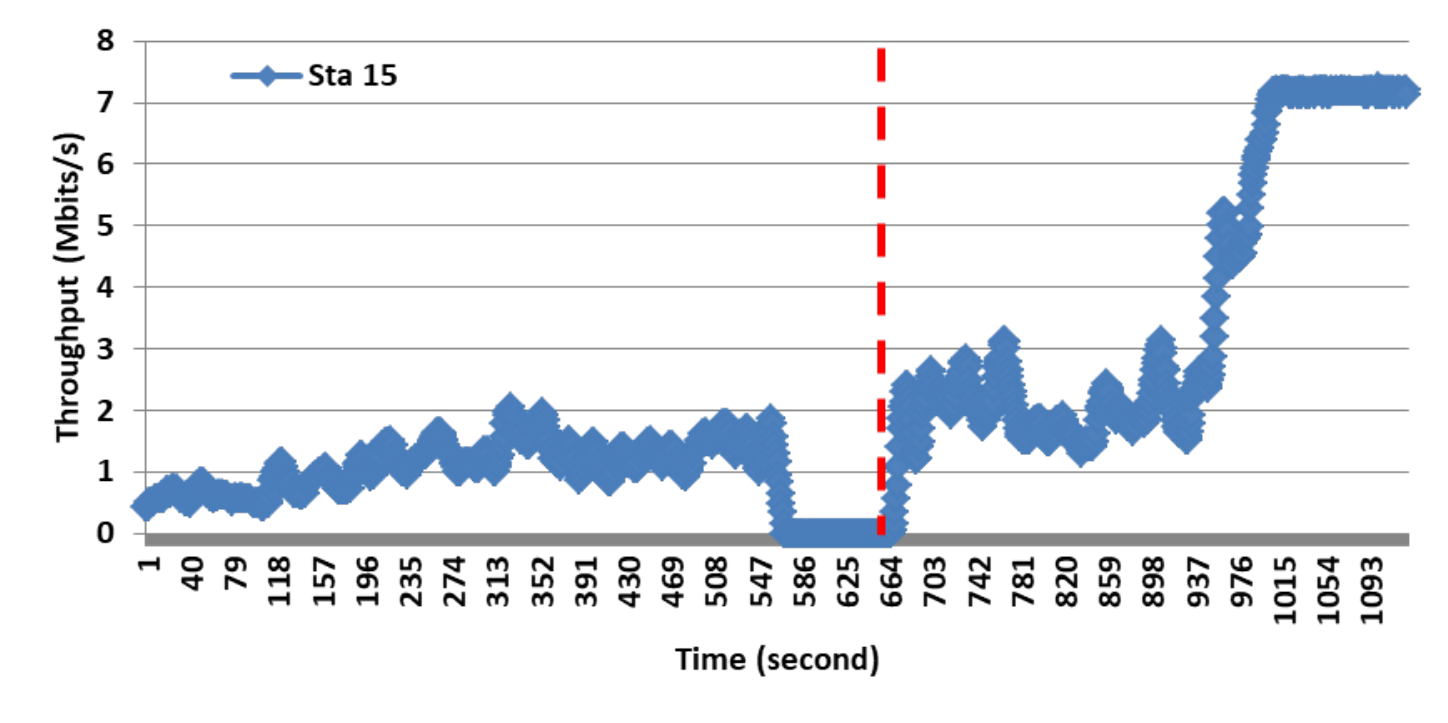}
}
\subfigure[Sta 16]{ \label{fig_throughtput_1:subfig:j}  
\includegraphics[width=2in,height=1.3in]{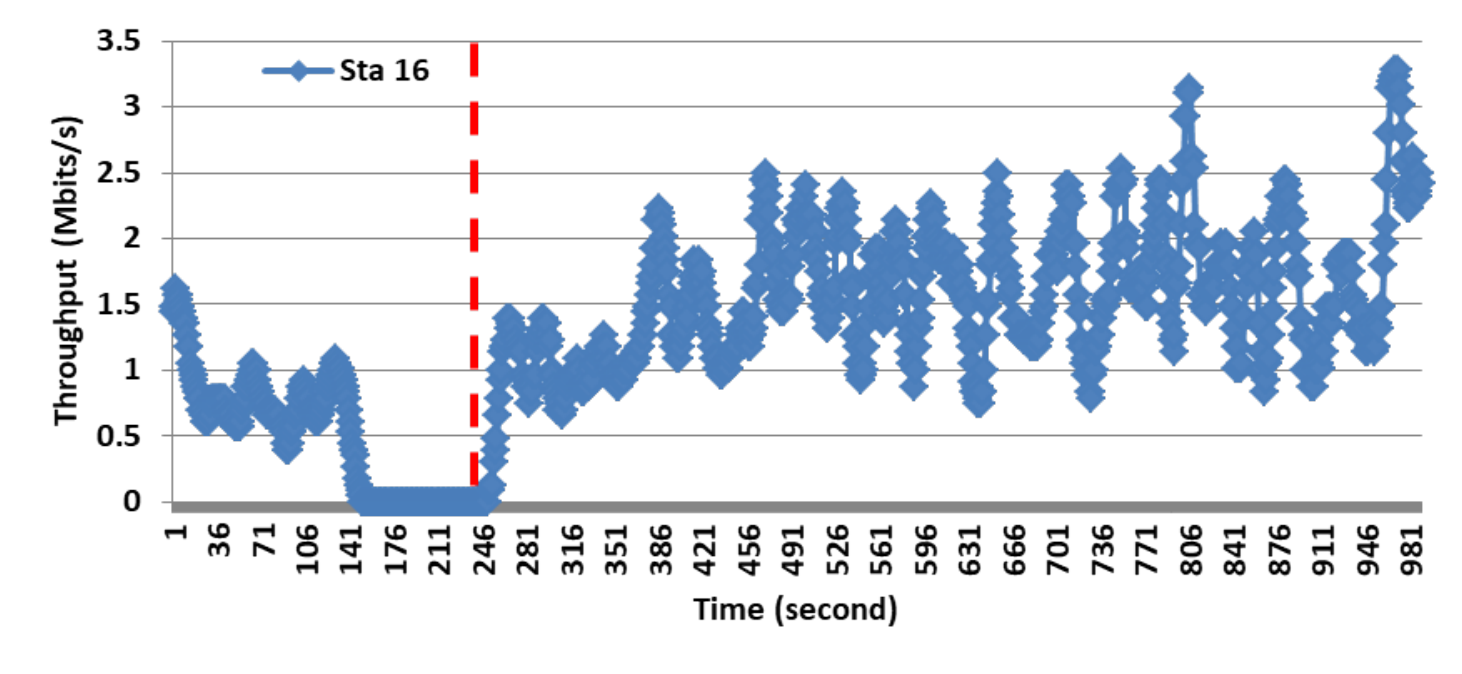}
}
\subfigure[Sta 17]{ \label{fig_throughtput_1:subfig:k}  
\includegraphics[width=2in,height=1.3in]{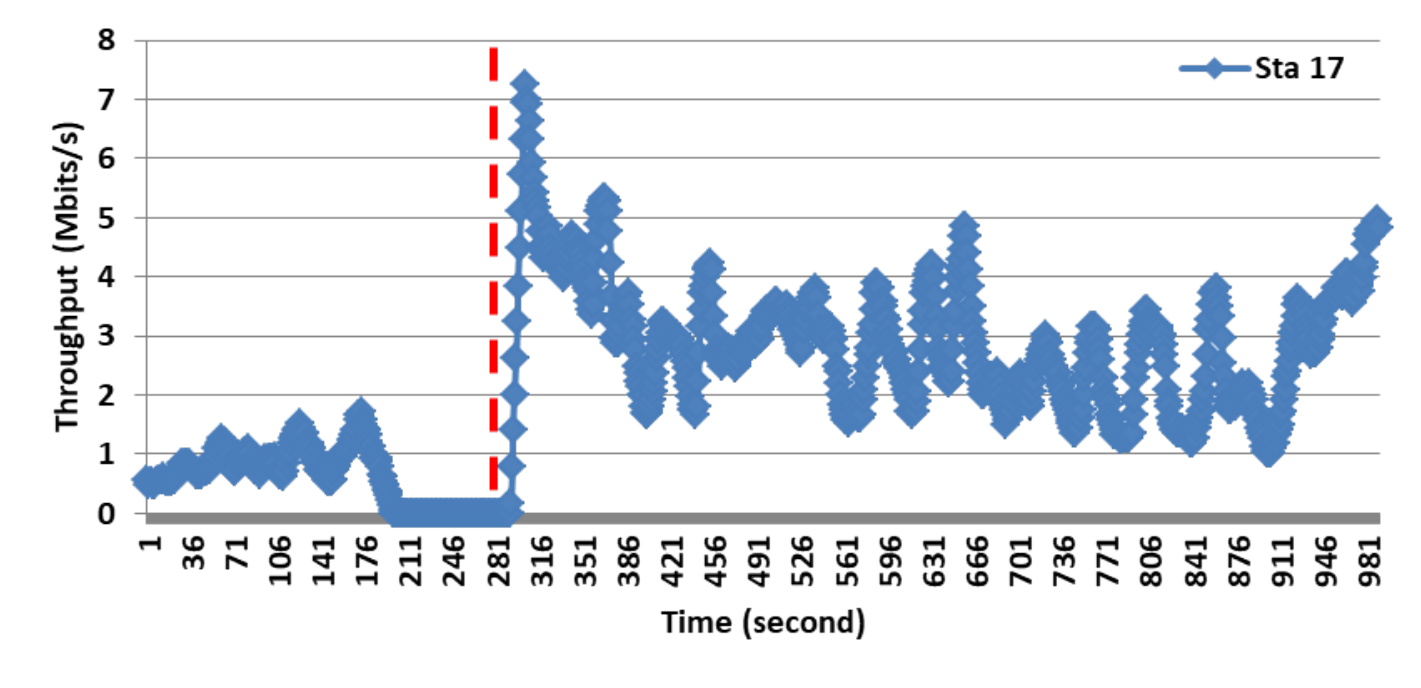}
}
\subfigure[Sta 18]{ \label{fig_throughtput_1:subfig:l}  
\includegraphics[width=2in,height=1.3in]{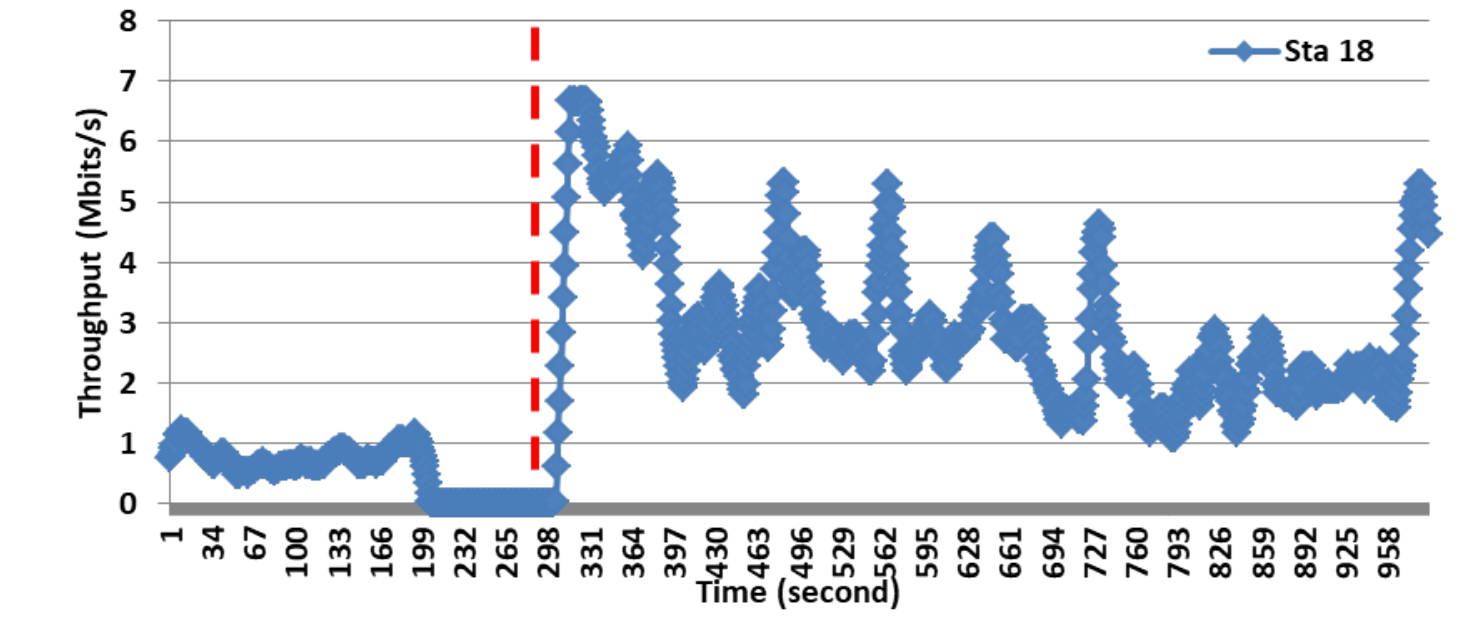}
}
\subfigure[Sta 19]{ \label{fig_throughtput_1:subfig:m}  
\includegraphics[width=2in,height=1.3in]{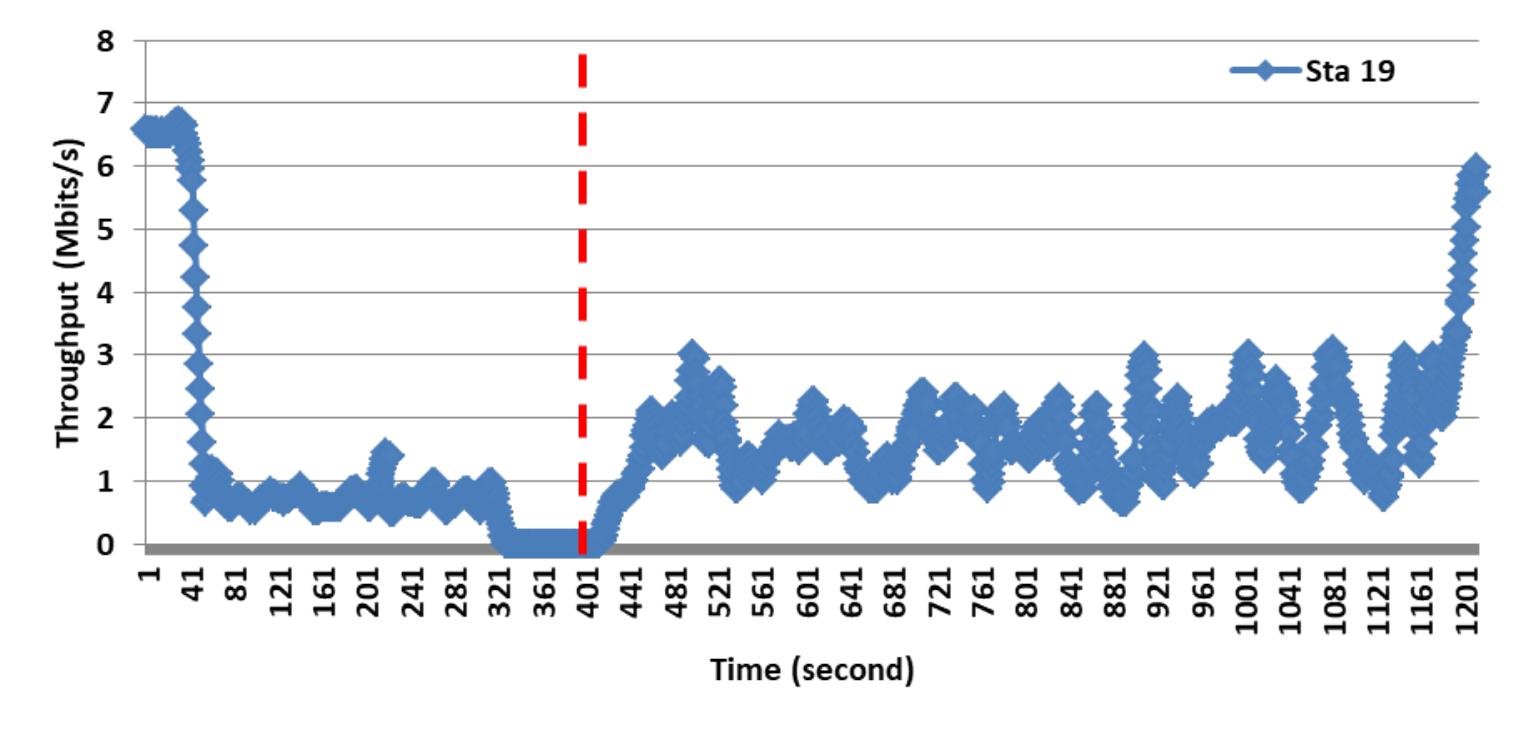}
}
\subfigure[Sta 20]{ \label{fig_throughtput_1:subfig:n}  
\includegraphics[width=2in,height=1.3in]{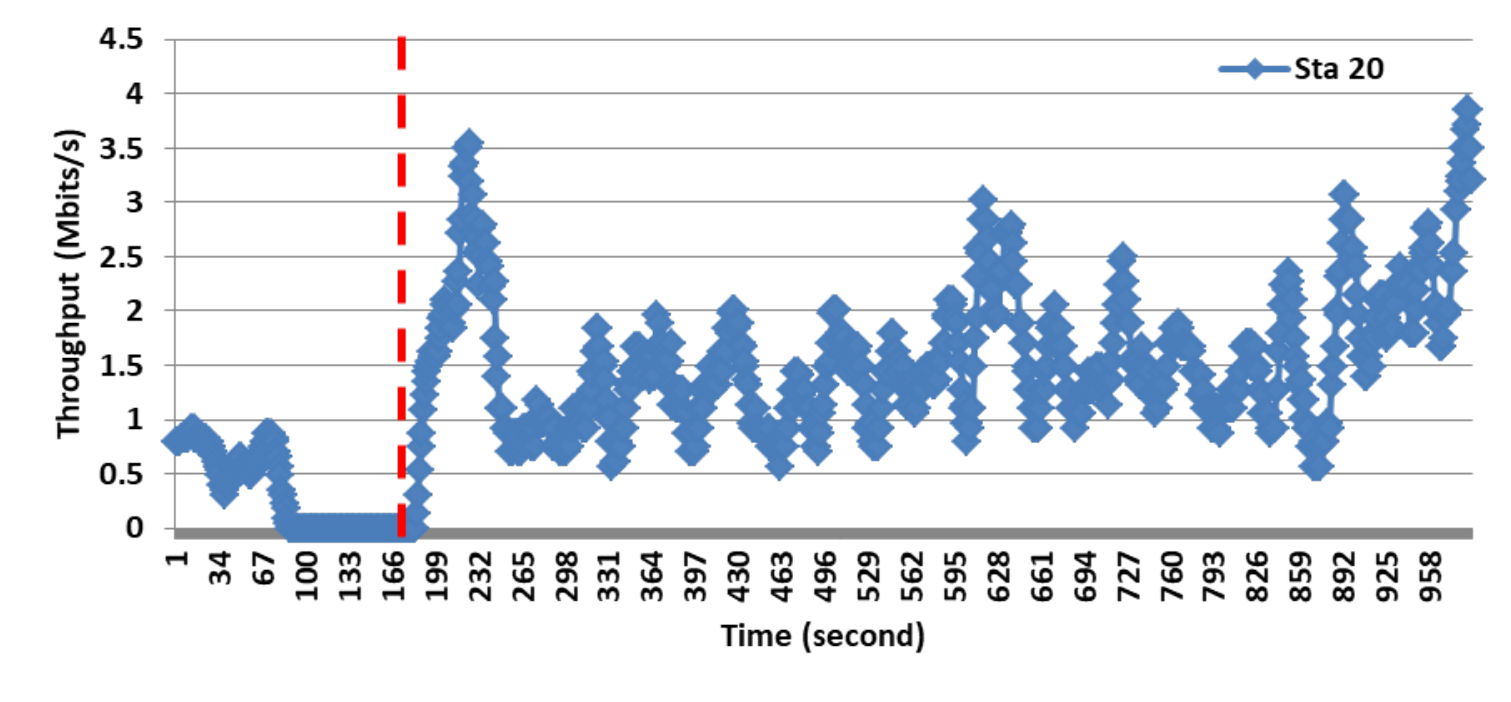}
}
\subfigure[Sta 21]{ \label{fig_throughtput_1:subfig:o}  
\includegraphics[width=2in,height=1.3in]{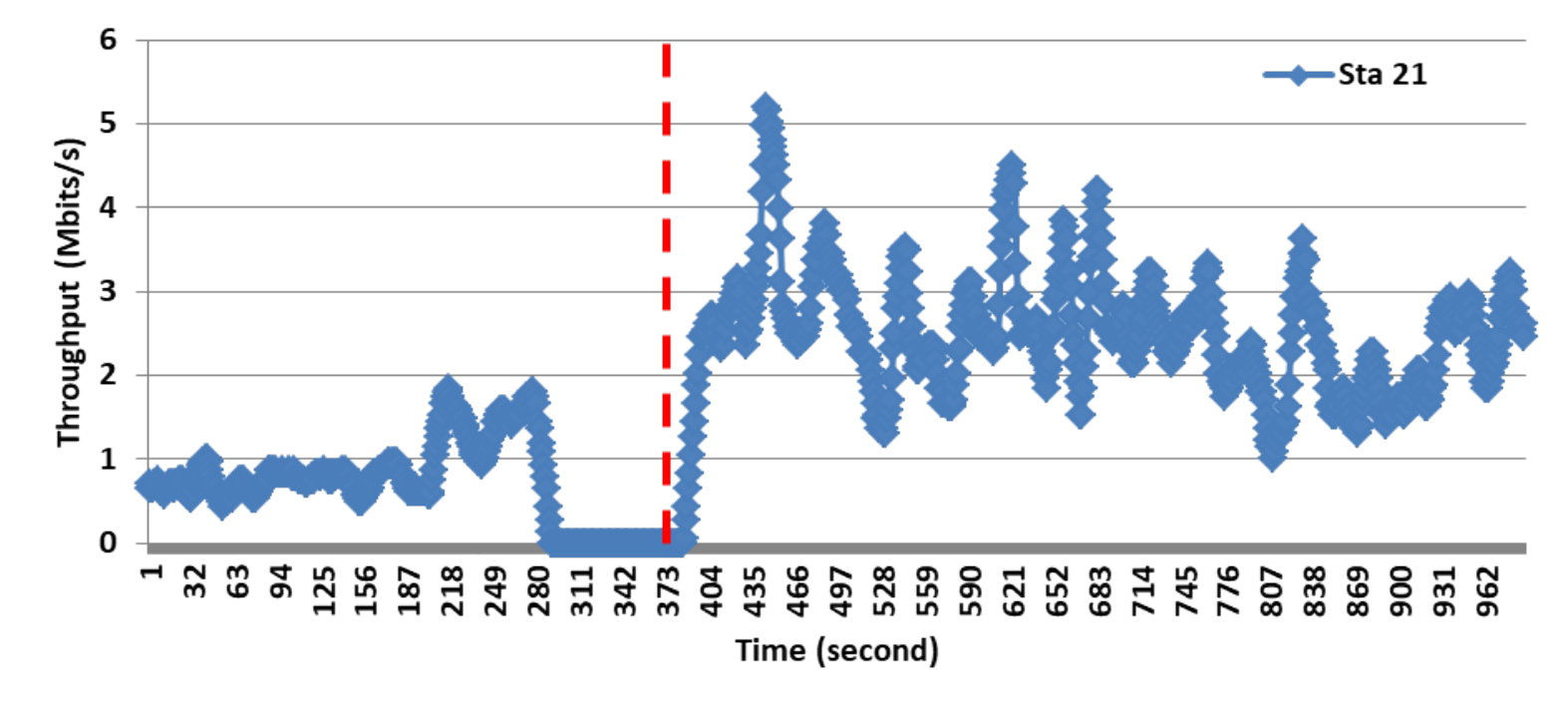}
}
\subfigure[Sta 26]{ \label{fig_throughtput_1:subfig:p}  
\includegraphics[width=2in,height=1.3in]{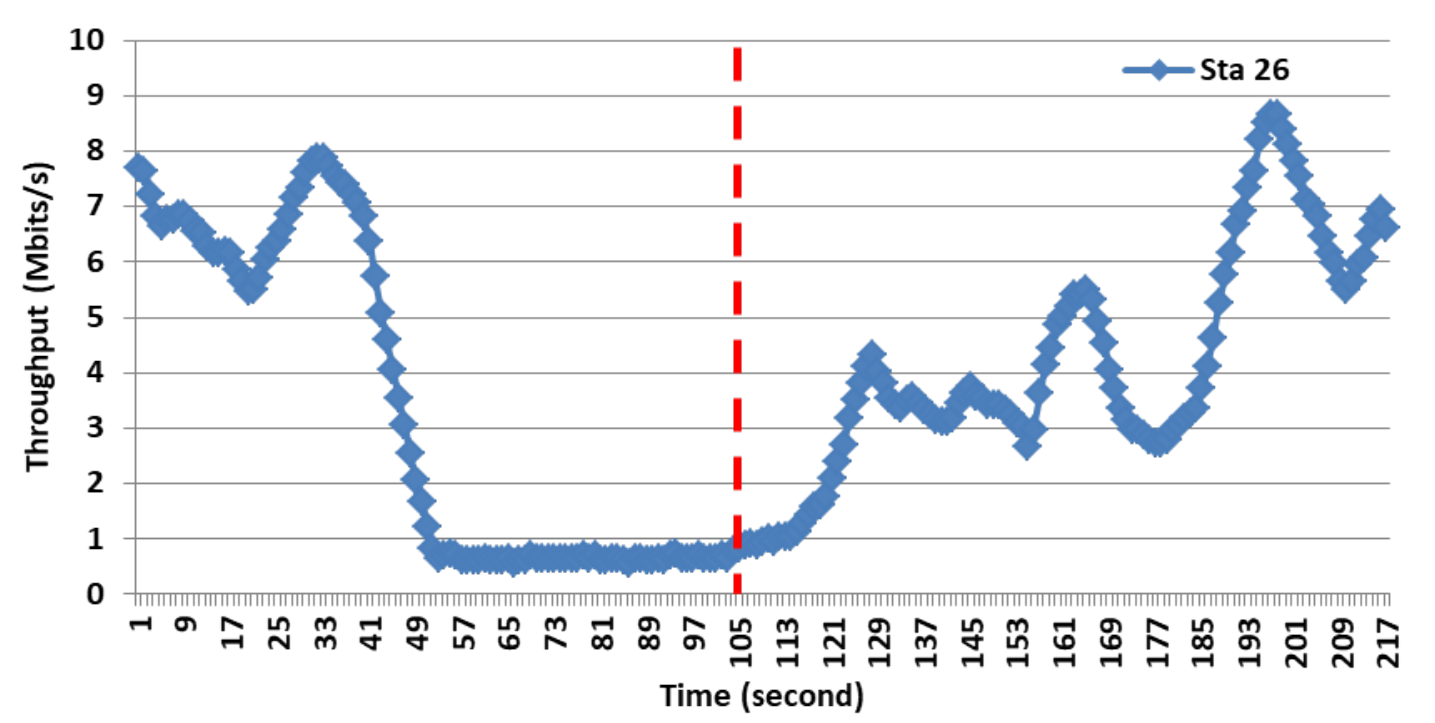}
}
\caption{Result end-to-end throughput of stations which the SARA algorithm is applied. The stations shown in the figure are Sta 1, 2, 3, 4, 6, 7, 10, 12, 15, 16, 17, 18, 19, 20, 21, 26. The red dash line in each figure denotes the time when SARA algorithm is executed.}
\label{fig_throughput}
\end{figure*}

\begin{figure*}[htbp]
\centering
\subfigure[Sta 5]{ \label{fig_throughtput_no:subfig:i}  
\includegraphics[width=2in,height=1.3in]{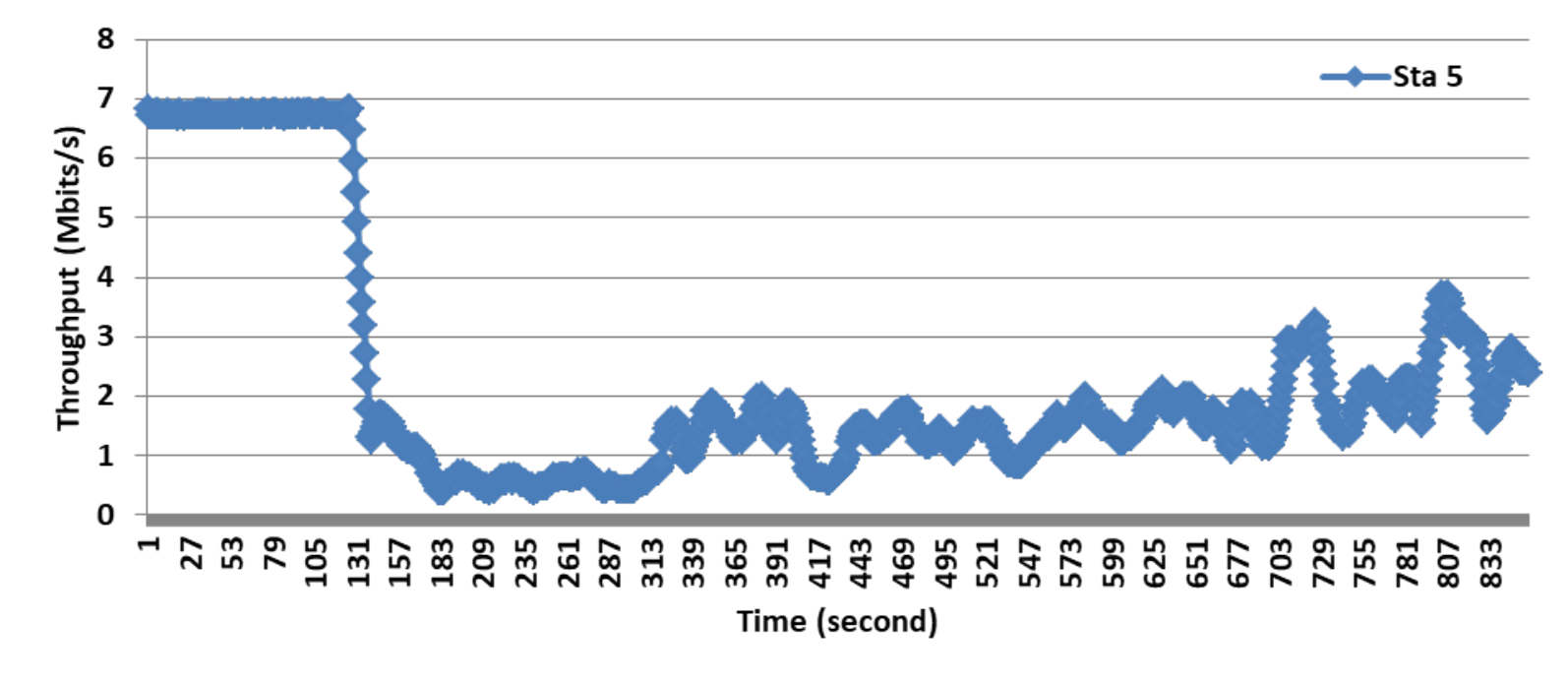}
}
\subfigure[Sta 8]{ \label{fig_throughtput_no:subfig:j}  
\includegraphics[width=2in,height=1.3in]{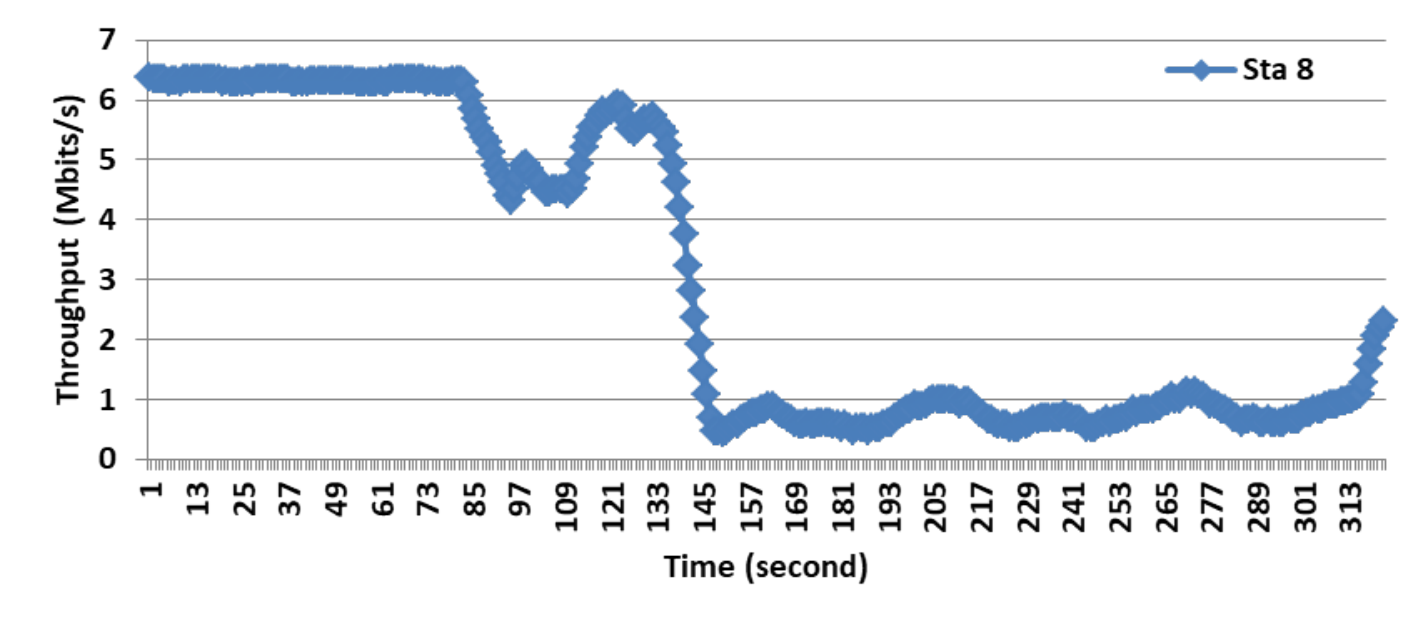}
}
\subfigure[Sta 9]{ \label{fig_throughtput_no:subfig:k}  
\includegraphics[width=2in,height=1.3in]{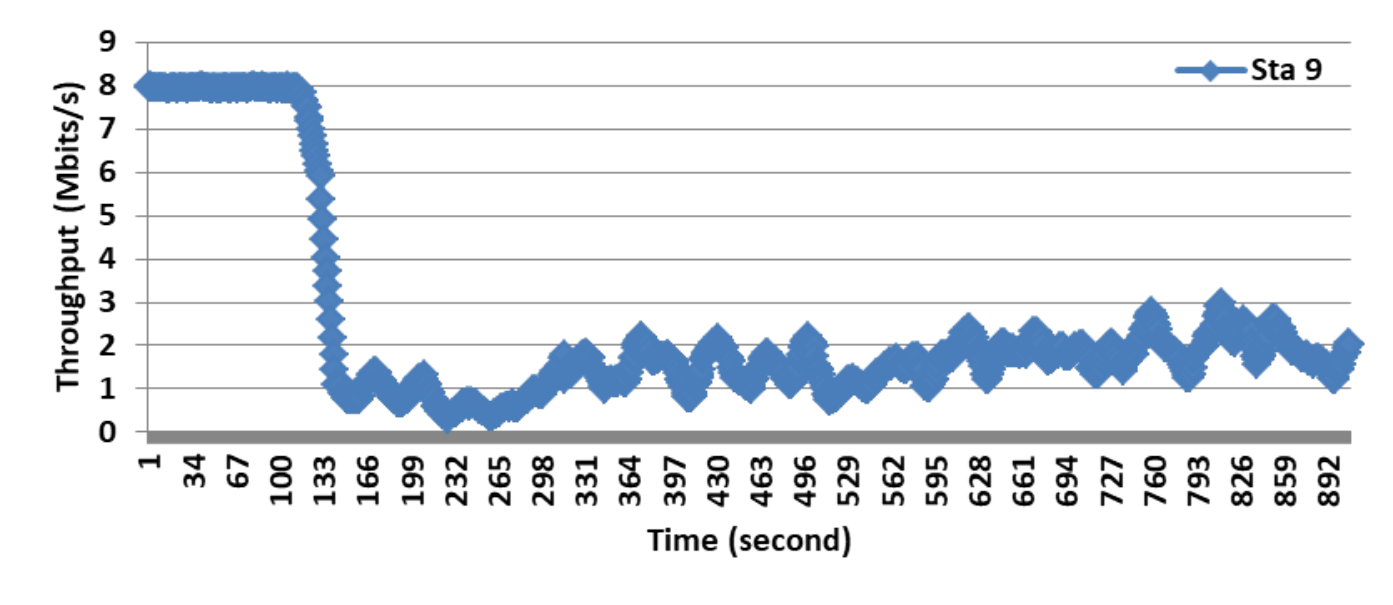}
}
\subfigure[Sta 11]{ \label{fig_throughtput_no:subfig:l}  
\includegraphics[width=2in,height=1.3in]{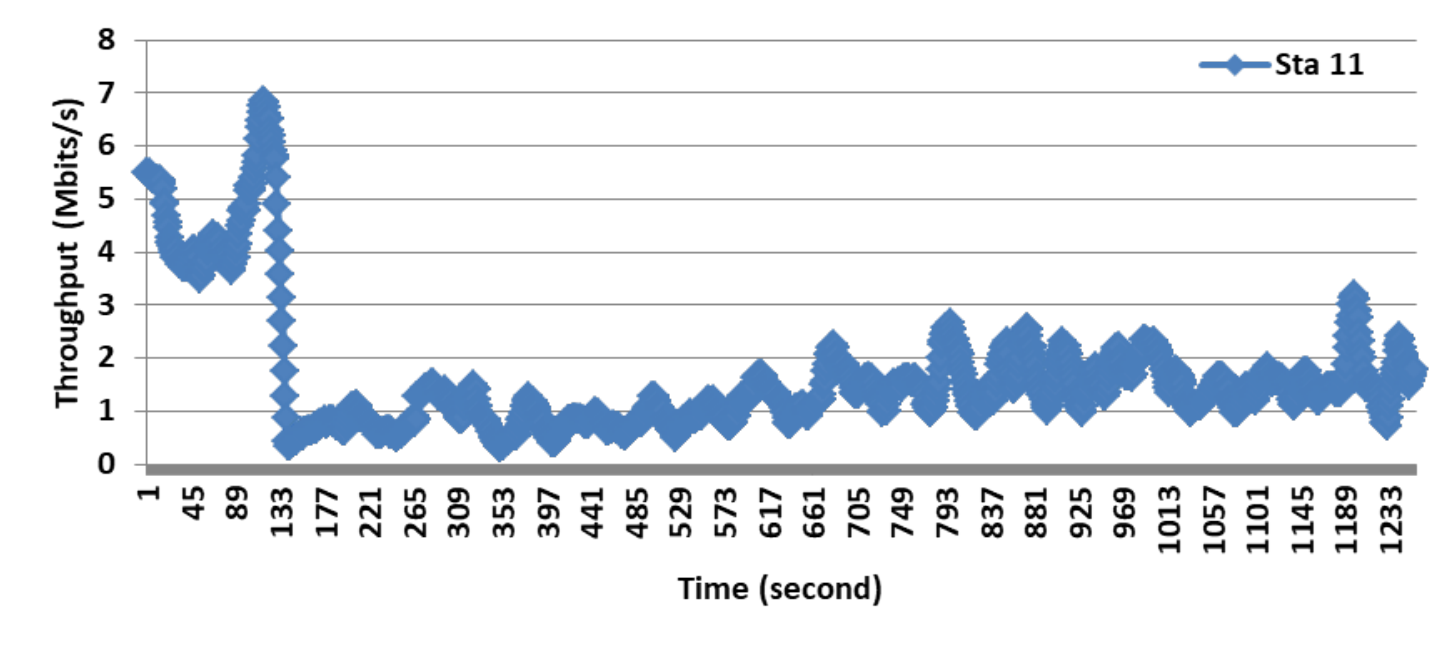}
}
\subfigure[Sta 13]{ \label{fig_throughtput_no:subfig:m}  
\includegraphics[width=2in,height=1.3in]{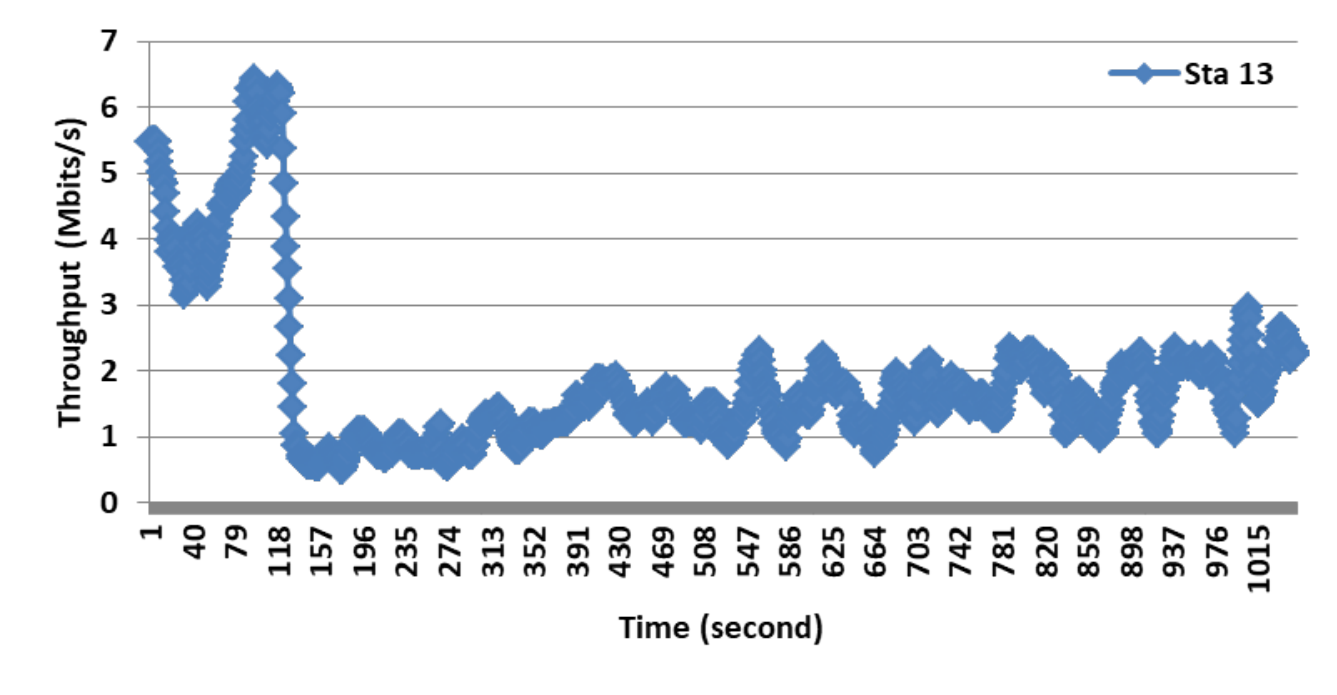}
}
\subfigure[Sta 27]{ \label{fig_throughtput_no:subfig:p}  
\includegraphics[width=2in,height=1.3in]{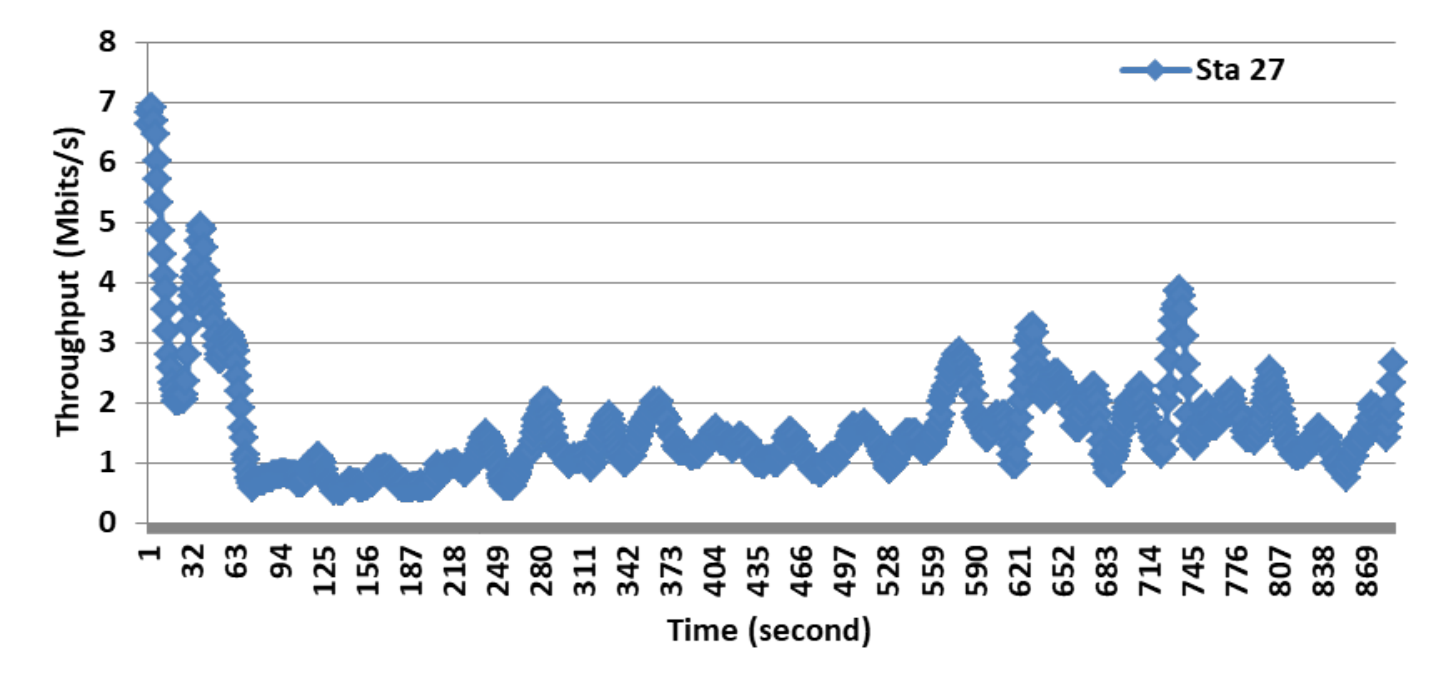}
}
\subfigure[Sta 14]{ \label{fig_throughtput_no_1:subfig:n}  
\includegraphics[width=2in,height=1.3in]{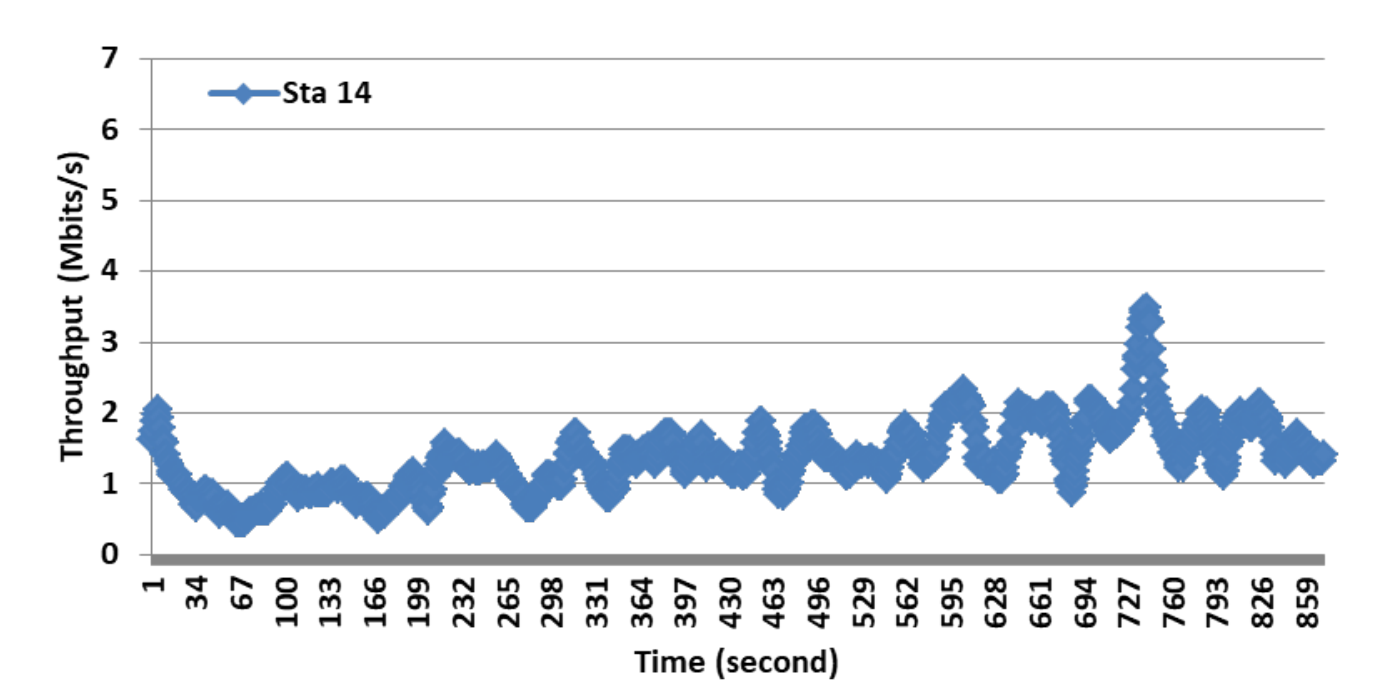}
}
\subfigure[Sta 22]{ \label{fig_throughtput_no_1:subfig:o}  
\includegraphics[width=2in,height=1.3in]{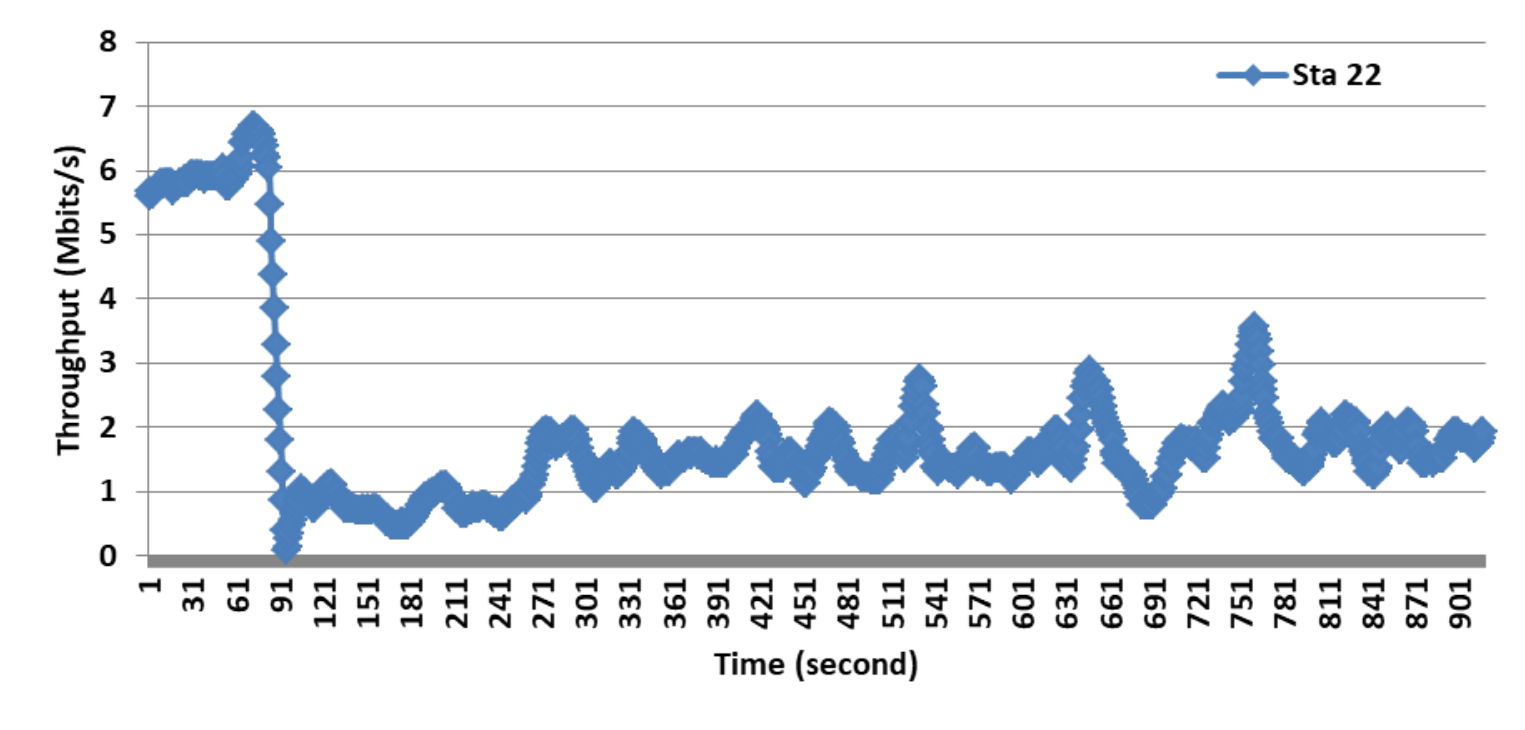}
}
\subfigure[Sta 23]{ \label{fig_throughtput_no_1:subfig:p}  
\includegraphics[width=2in,height=1.3in]{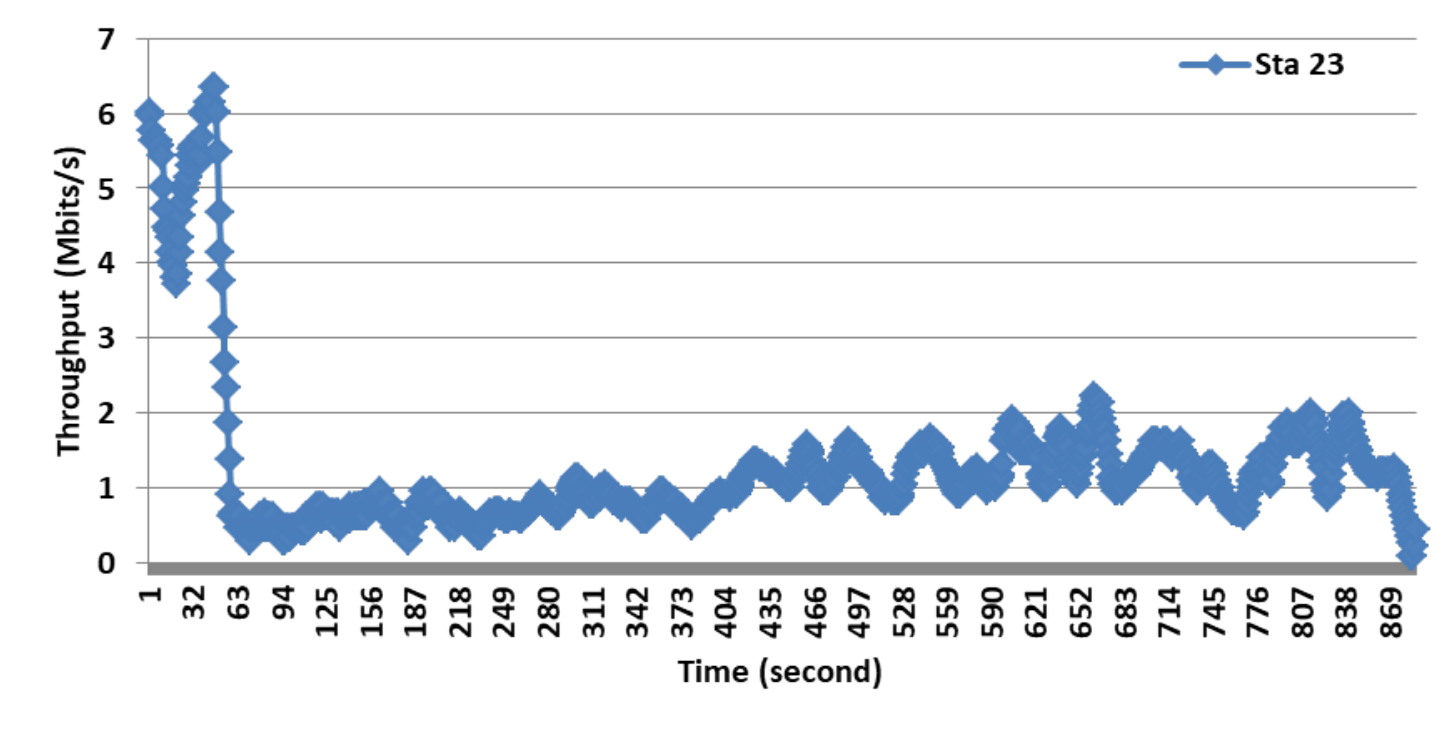}
}
\subfigure[Sta 24]{ \label{fig_throughtput_no_1:subfig:p}  
\includegraphics[width=2in,height=1.3in]{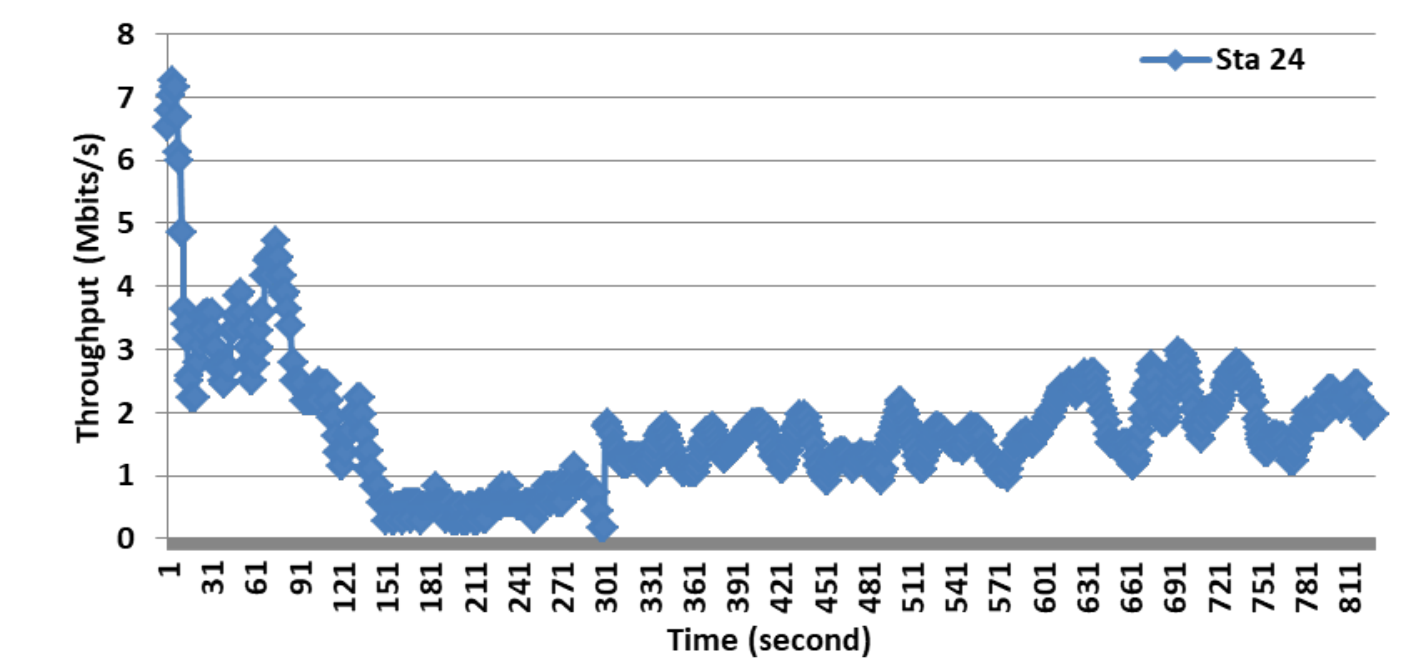}
}
\subfigure[Sta 25]{ \label{fig_throughtput_no_1:subfig:p}  
\includegraphics[width=2in,height=1.3in]{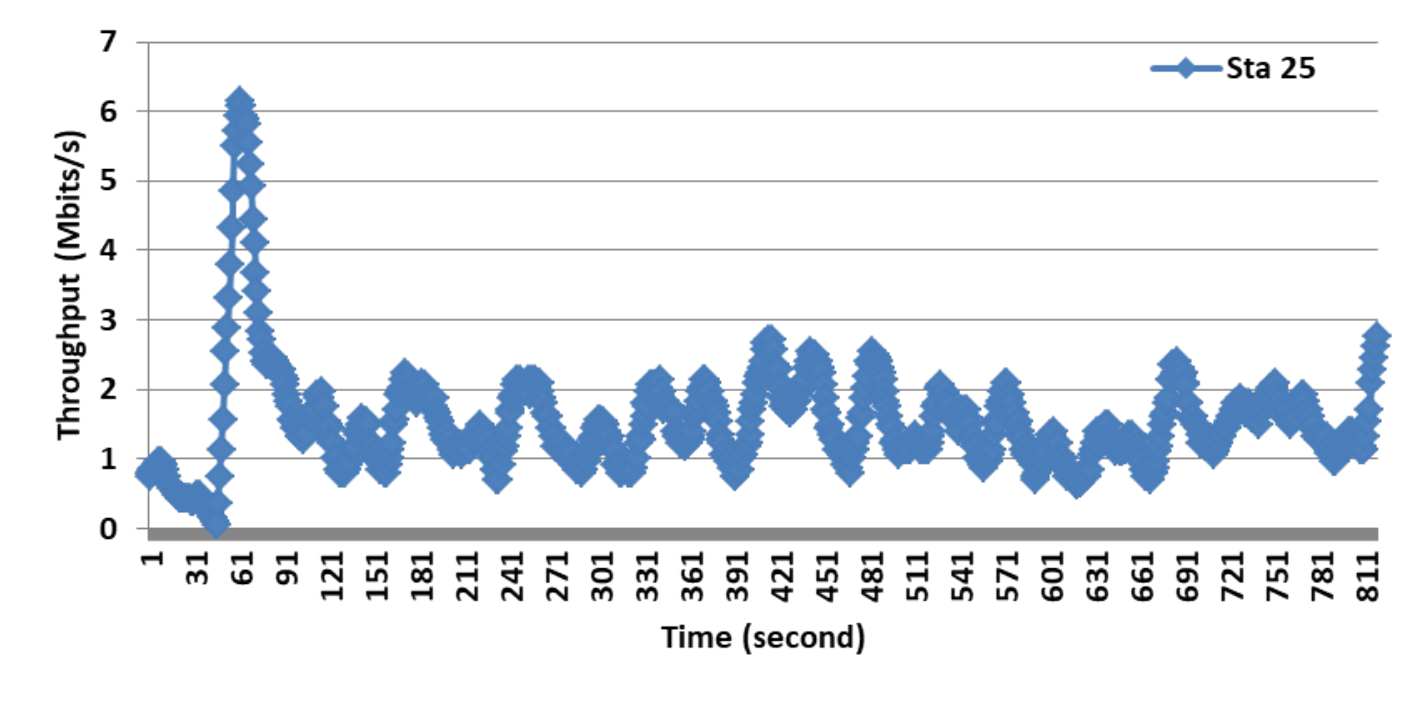}
}
\caption{Result end-to-end throughput of stations which the SSS algorithm is applied. The stations shown in the figure are Sta 5, 8, 9, 11, 13, 14, 22, 23, 24, 25, 27.}
\label{fig_throughput_no}
\end{figure*}

\begin{figure}[hbt]
\centering
\includegraphics[width=0.5\textwidth]{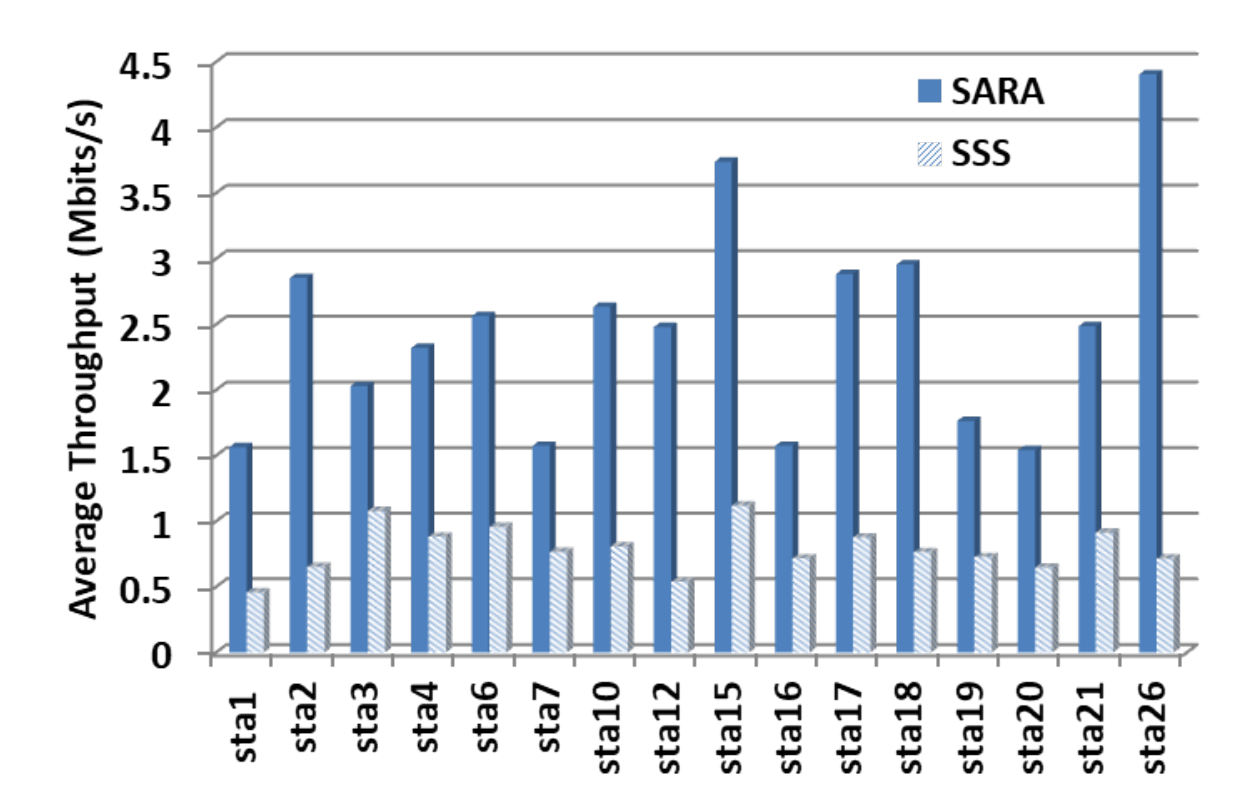}
\caption{The comparison of the two AP selection methods, SSS and SARA, in terms of throughput, on the 16 mobile stations which both methods have been applied. }
\label{fig_compare} 
\end{figure}

\begin{figure}[hbt]
\centering
\includegraphics[width=0.33\textwidth]{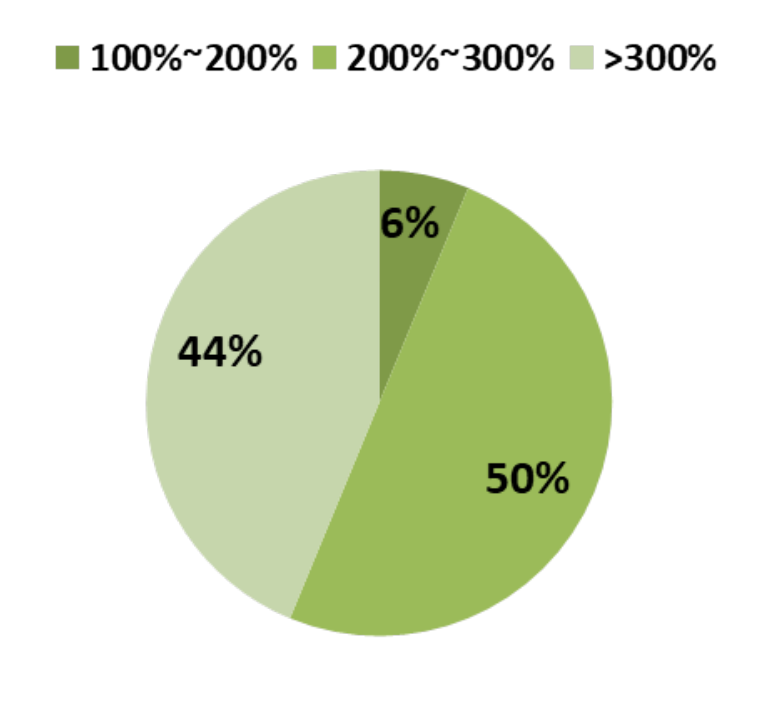}
\caption{The statistics of the throughput gain of all the mobile stations executed SARA.}
\label{fig_gain_stat} 
\end{figure}

\begin{figure}[hbt]
\centering
\includegraphics[width=0.39\textwidth]{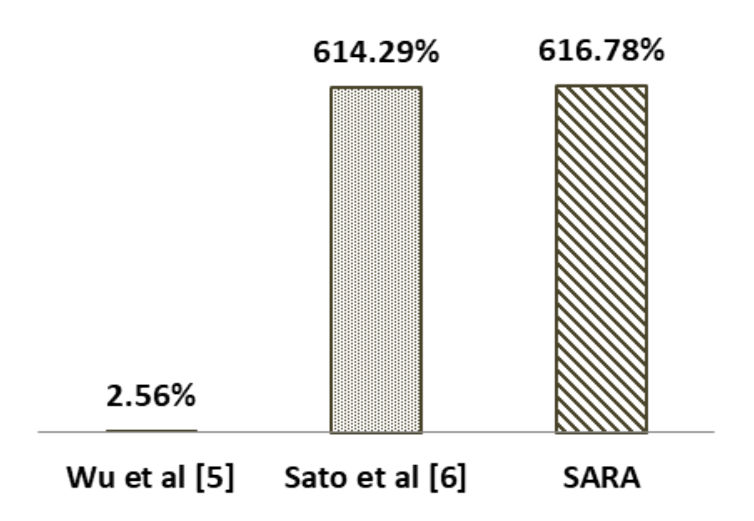}
\caption{The comparison of the best case bandwidth gain achieved after hybrid AP selection methods are performed, between our proposed method SARA and two method from related work \cite{wu2016two,sato2015resilient}. }
\label{fig_compare_gain}
\end{figure}

\begin{table}[ht]
\centering
\caption{Throughput Gain Experienced After Executing SARA}
\label{table_gain}
\resizebox{0.37\textwidth}{35mm}{
\begin{tabular}{|c|c|}
\hline
Mobile Station & Throughput Gain (\%) \\ \hline
Sta 1 & $214.50\%$ \\
Sta 2 & $437.93\%$ \\
Sta 3 & $188.50\%$ \\
Sta 4 & $263.19\%$ \\
Sta 6 & $ 268.40\% $ \\
Sta 7 & $ 205.80\% $ \\
Sta 10 & $ 326.68\% $ \\
Sta 12 & $ 461.30\% $ \\
Sta 15 & $ 334.10\% $ \\
Sta 16 & $ 219.96\% $ \\
Sta 17 & $ 329.51\% $ \\
Sta 18 & $ 386.96\% $ \\
Sta 19 & $ 243.12\% $ \\
Sta 20 & $ 238.42\% $ \\
Sta 21 & $ 272.39\% $ \\
Sta 26 & $ 616.78\% $ \\
\hline
\end{tabular}}
\end{table}

\begin{table*}[ht]
\tiny
\centering
\caption{The Comparison with Other Hybrid AP Selection Methods}
\label{table_compare_hybrid}
\resizebox{\textwidth}{9mm}{
\begin{tabular}{|c|cccc|}
\hline
Method & Hybrid Technologies & Theory & Simulation Environment & Best Throughput Gain \\
&&&&Comparing to SSS \\ \hline
Proposed by Wu et al \cite{wu2016two} & LiFi, WiFi & Fuzzy Logic & simulation with Matlab & 2.56\% \\
Proposed by Sato et al \cite{sato2015resilient} & LTE, WiFi, Satellite, FTTH & AHP & Real Testbed & 614.29\% \\
SARA & WiFi, LTE, Satellite & Knowledge Graph Based & emulation on Mininet-WiFi & 616.78\% \\
\hline
\end{tabular}}
\end{table*}

\begin{figure}[hbt]
\centering
\includegraphics[width=0.43\textwidth]{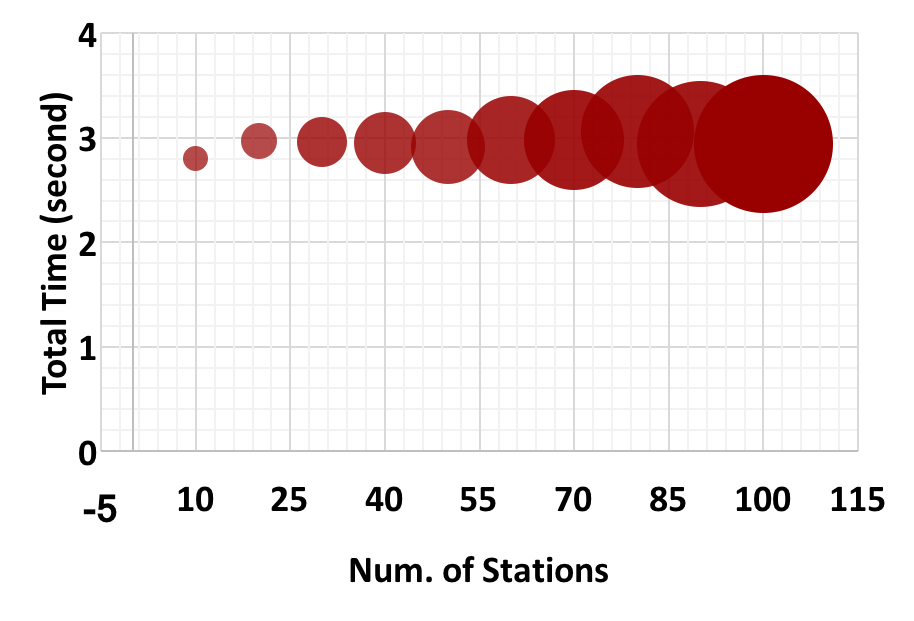}
 \caption{Total time to complete one AP selection circle: time to 1) scan all available APs, 2) test all available APs, 3) sort the APs and choose the best one. The experiment is repeated 10 times on networks with different number of stations. Each dot represents the system time overhead in one experiment. The diameter of a dot represent  the size of the knowledge base for each network (the number of RDF triples inside the knowledge base). As can be seen from the figure, the total execution time is quite stable with the increase of network scale, at roughly $3~seconds$, which is acceptable for an autonomic AP selection service.}
 \label{fig_system} 
\end{figure}

\section{Evaluation}
In this section, we describe the experimental results of the current AP selection method and the proposed method. We implement all the methods in an emulated Mininte-WiFi environment. 

We emulate a network with hybrid access technologies of WiFi, satellite, LTE, which includs 9 WiFi APs, 1 satellite AP, 1 LTE AP and 27 stations in a 2D square area of $300m\times300m$, as shown in Fig. \ref{fig_move}. The parameters of the emulated network are shown in Table \ref{table_para}. 

To simplify the scenario, all the stations are connected with the server \emph{``h1''} via WiFi APs following the SSS AP selection rule by default. The server \emph{``h1''} is initiated processing all the video, audio, and text uploaded by all the stations.

The simulation is executed using the following steps.

\begin{itemize}[leftmargin=0em]
\setlength{\itemindent}{0.3in}
\item[1 --] 27 Mobile stations are generated, distributing randomly allover the area, wandering at random direction with random speed. By default, the mobile stations are handovered among IEEE 802.11b APs according to SSS method. An \emph{iperf} tool is started on each one of the stations from the beginning of the experiment, testing the bandwidth between themselves and the server \emph{``h1''} in real time through out the experiment. The bandwidth data are recorded as the results for analysis.

\item[2 --] After 400 second, an incident happens at location (20, 20), and all the mobile stations began to gather (at different reaction time ranging from 401 to 413 seconds) toward the incident location. The distribution of mobile stations at different time point can be seen in Fig. \ref{fig_move}. 

\item[3 --] After the movement, all the mobile stations are settled in an rectangle area \{(5,5), (30, 30)\} near the incident location. At randomly chosen time, the SARA is executed on randomly chosen mobile stations. In this experiment, 16 mobile stations, out of the total 27 mobile stations are chosen to execute SARA. The other mobile stations will keep using SSS AP selection method.
\end{itemize}

\subsection{Experiment Environment}
The evaluation experiment is carried out on Mininet-WiFi \cite{fontes2015mininet}, which is a wireless network emulator. The evaluation was carried out on an virtual machine running on 64 bit Ubuntu 14.0.4, with 1024 MB base memory, and signal CPU. The virtual storage is 8 GB, while the actual storage is 3.28 GB.
The host machine is a MacBook Air OS X 10.9.5 with 3 MB cache running on Intel Core i5 at 1.5 GHz. The system has 128 GB SSD and 4 GB RAM. 

The following tools were adopted to build and query the semantic knowledge base.

\begin{itemize}[leftmargin=0em]
\setlength{\itemindent}{0.3in}
\item{\textbf{\emph{Rdflib:}}} version 4.2.1. A Python library to work with RDF;
\item{\textbf{\emph{Neo4j:}}} version 1.0.2. A graph database storage and process engine. 
\end{itemize}

The tools adopted to monitor the network throughput were:

\begin{itemize}[leftmargin=0em]
\setlength{\itemindent}{0.3in}
\item{\textbf{\emph{iperf:}}} measure the available channel bandwidth;
\item{\textbf{\emph{iwconfig:}}} to configure a wireless network interface;
\item{\textbf{\emph{iw:}}} to show and/or monitor the wireless network devices and their configuration;
\item{\textbf{\emph{ping:}}} to check the connectivity of a connection and test the round-trip time.
\end{itemize}

\subsection{Evaluation Results in Throughput}
When evaluating our proposed implementation SARA, we tried to seek answers for the following questions:

\begin{itemize}[leftmargin=0em]
\setlength{\itemindent}{0.3in}
	\item[-] \emph{Can the SARA successfully reselect and reconnect an AP for the mobile stations, based on given rules?}
	\item[-] \emph{How much does user's channel connection (in terms of throughput) improve after SARA is performed on the mobile station?}
	\item[-] \emph{What is the overhead in terms of switching time? Is it acceptable?}
\end{itemize}

\subsubsection{Comparison with Current AP selection Method -- ``SSS''}
The throughput results for all the mobile stations with and without SARA performing on it are shown in Fig. \ref{fig_throughput} and \ref{fig_throughput_no}, respectively. In Fig. \ref{fig_throughput} and Fig. \ref{fig_throughput_no}, the x-axis is the time elapsed since the beginning of the experiment, and the y-axis is the end-to-end throughput detected by \emph{iperf}. In Fig. \ref{fig_throughput}, the time when SARA is executed is denoted by a red dash line in each subfigures. As the mobile stations are randomly chosen at random time to execute SARA, the positions of the red dash lines in each subfigure are different. 

As can be observed from both Fig. \ref{fig_throughput} and Fig. \ref{fig_throughput_no}, soon after the incident start, and the mobile station begin to move toward the incident location, the channel connection deteriorate significantly, although the deteriorate happen at different time, for the location and movement of each mobile station are random. In Fig. \ref{fig_throughput}, after SARA is executed, a significant raise can be observed in every mobile station, while in Fig. \ref{fig_throughput_no}, where the current AP selection method SSS is adopted, the throughput stays the same and does not experience obvious increase. The effect of SARA on channel condition improvement is evident. 

The comparison between current AP selection method SSS and SARA is further investigated in terms of the throughput on the same mobile station, as presented in a bar chart in Fig. \ref{fig_compare}. The x-axis in Fig. \ref{fig_compare} denotes each mobile station where both SSS and SARA are executed. The dark blue bar is the average of the throughput results measured when SARA is executed, while the bar in pale blue is the average throughput value when SSS is executed. It is obviously shown in Fig. \ref{fig_compare} that SARA is able to double the throughput at the worst case, on all the mobile stations. At the best case, as shown at the Sta 26, the throughput is increased more than 6 times than it uses SSS method.

The exact throughput gain experienced by each mobile station is given in Table \ref{table_gain}. A statistic of these results is presented as a pie chart in Fig. \ref{fig_gain_stat}. As shown in the Table \ref{table_gain} and Fig. \ref{fig_gain_stat}, almost all the mobile stations (96\%) have their throughput more than doubled after SARA is executed. Half of the mobile stations have tripled their throughput after executing SARA.

\subsubsection{Comparison with Other Hybrid AP Selection Methods}
In this section, we compared SARA with two different approaches for hybrid AP selection proposed by Wu et al \cite{wu2016two} and Sato et al \cite{sato2015resilient} respectively. These approaches are compared in Table \ref{table_compare_hybrid}, in terms of the access technologies consisted in the hybrid network, the backup theory adopted, the experiment environment, and the result throughput gain comparing to SSS. As shown in the table, different mathematical theories are adopted to reselect AP in a hybrid network. Sato et al \cite{sato2015resilient} use real testbed consisting of several network interface cards (NICs) and Linux based computers. The other methods are only evaluated on simulated environment.  

The performance of each method in terms of the best case throughput gain (comparing to the current AP selection method SSS) is presented as a bar chart in Fig. \ref{fig_compare_gain}. It is evident that our proposed system SARA achieves the highest throughput gain among the three. Sato et al's system performs slightly poorer, but considering the limited emulation environment SARA is implemented (a virtual machine with only 3.28 GB actual storage and 1024 MB base memory, and the bandwidth of the emulated link by Mininet-WiFi is restricted to no more than 1 Mbits/s), and the professional devices Sato et al adopted, it stands a good chance that the advantage of SARA be more significant if applied in real testbed. Another method evaluated by simulation, Wu et al \cite{wu2016two}, achieves only a slightly advantage of 2.56\%, comparing to the current method SSS. Thus the efficiency of SARA is evident.

\subsection{Overhead}
We investigate the overhead of SARA in terms of execution time. This execution time is the time from the start of executing SARA until the AP re-selection is complete. A diverse set of time overhead on the 10 networks with the number of mobile stations varying from 10 to 100 is collected. The time overheads of networks with different scale and different knowledge graph are illustrated in Fig. \ref{fig_system}. In Fig. \ref{fig_system}, the time overhead value is denoted on the y-axis, the scale of the network is denoted in the x-axis, and the width of the dots denotes the scale of the semantic knowledge graph adopted by each network. As expected, the scale of knowledge base is proportional to the scale of the network. Thus, the width of dots grows with the x-axis. It can be shown in Fig. \ref{fig_system} that the system time is roughly 3s, which is more efficient comparing the traditional AP selection strategy (more than 10 seconds)\cite{nicholson2006improved}, and there is no obvious increase as the network scale increases. This is probably due to the fact that SARA is an application on a knowledge-based network management system (SEANET), and thus the execution time depend heavier on the speed of querying the knowledge base than the scale and complexity of the network. Thus, the knowledge-based system not only enabled an autonomic fashion of network management, but also make it highly efficient. In some instances, it resulted in the execution time being slightly less than the network with smaller number of nodes. This can be observed from the time overhead with the number of station $=50$ illustrated in Fig. \ref{fig_system}.

We argue that even with the above time overhead, SARA could provide more reasonable AP selection by taking the real QoS into consideration and it is accomplished in an automatic fashion.

\section{Conclusions}
In current wireless telecommunications networks, a combination of various access technologies is available for mobile users, e.g., WiFi, LTE, 3G/2G, Satellite, LiFi, mesh/adhoc. Users need to choose between these technologies to achieve the best communication quality. Current AP selection criteria is based only on signal strength 
\par
We have presented SARA, an autonomic resource allocation service for hybrid wireless networks, as an application of SEANET system. Based on knowledge base built by SEANET, SARA can quickly associate with the selected AP taking account of the bandwidth, the congestion, and the signal strength. 
\par
We have evaluated SARA in an incident scenario on a emulated hybrid wireless network. We argue that the evaluation results has proven the accuracy and performance of SARA. Our overhead is acceptable.


\bibliographystyle{IEEEtran} 
\bibliography{res}

\begin{thebibliography}{10}
\providecommand{\url}[1]{#1}
\csname url@samestyle\endcsname
\providecommand{\newblock}{\relax}
\providecommand{\bibinfo}[2]{#2}
\providecommand{\BIBentrySTDinterwordspacing}{\spaceskip=0pt\relax}
\providecommand{\BIBentryALTinterwordstretchfactor}{4}
\providecommand{\BIBentryALTinterwordspacing}{\spaceskip=\fontdimen2\font plus
\BIBentryALTinterwordstretchfactor\fontdimen3\font minus
  \fontdimen4\font\relax}
\providecommand{\BIBforeignlanguage}[2]{{%
\expandafter\ifx\csname l@#1\endcsname\relax
\typeout{** WARNING: IEEEtran.bst: No hyphenation pattern has been}%
\typeout{** loaded for the language `#1'. Using the pattern for}%
\typeout{** the default language instead.}%
\else
\language=\csname l@#1\endcsname
\fi
#2}}
\providecommand{\BIBdecl}{\relax}
\BIBdecl

\bibitem{nicholson2006improved}
A.~J. Nicholson, Y.~Chawathe, M.~Y. Chen, B.~D. Noble, and D.~Wetherall,
  ``Improved access point selection,'' in \emph{Proceedings of the 4th
  international conference on Mobile systems, applications and services}.\hskip
  1em plus 0.5em minus 0.4em\relax ACM, 2006, pp. 233--245.

\bibitem{vasudevan2005facilitating}
S.~Vasudevan, K.~Papagiannaki, C.~Diot, J.~Kurose, and D.~Towsley,
  ``Facilitating access point selection in ieee 802.11 wireless networks,'' in
  \emph{Proceedings of the 5th ACM SIGCOMM conference on Internet
  Measurement}.\hskip 1em plus 0.5em minus 0.4em\relax Usenix Association,
  2005, pp. 26--26.

\bibitem{fukuda2004decentralized}
Y.~Fukuda, T.~Abe, and Y.~Oie, ``Decentralized access point selection
  architecture for wireless lans,'' in \emph{Wireless Telecommunications
  Symposium}, vol. 2004, 2004, pp. 137--145.

\bibitem{wu2016two}
X.~Wu, D.~Basnayaka, M.~Safari, and H.~Haas, ``Two-stage access point selection
  for hybrid vlc and rf networks,'' in \emph{Personal, Indoor, and Mobile Radio
  Communications (PIMRC), 2016 IEEE 27th Annual International Symposium
  on}.\hskip 1em plus 0.5em minus 0.4em\relax IEEE, 2016, pp. 1--6.

\bibitem{sato2015resilient}
G.~Sato, N.~Uchida, and Y.~Shibata, ``Resilient disaster network based on
  software defined cognitive wireless network technology,'' \emph{Mobile
  Information Systems}, vol. 2015, 2015.

\bibitem{chen2013access}
X.~Chen, W.~Yuan, W.~Cheng, W.~Liu, and H.~Leung, ``Access point selection
  under qos requirements in variable channel-width wlans,'' \emph{IEEE Wireless
  Communications Letters}, vol.~2, no.~1, pp. 114--117, 2013.

\bibitem{westerinen2001terminology}
A.~Westerinen, J.~Schnizlein, J.~Strassner, M.~Scherling, B.~Quinn, S.~Herzog,
  A.~Huynh, M.~Carlson, J.~Perry, and S.~Waldbusser, ``Terminology for
  policy-based management,'' Tech. Rep., 2001.

\bibitem{houidi2016knowledge}
Z.~B. Houidi, ``A knowledge-based systems approach to reason about
  networking,'' in \emph{Proceedings of the 15th ACM Workshop on Hot Topics in
  Networks}.\hskip 1em plus 0.5em minus 0.4em\relax ACM, 2016, pp. 22--28.

\bibitem{zhou2015network}
Q.~Zhou, C.-X. Wang, S.~McLaughlin, and X.~Zhou, ``Network virtualization and
  resource description in software-defined wireless networks,'' \emph{IEEE
  Communications Magazine}, vol.~53, no.~11, pp. 110--117, 2015.

\bibitem{180}
S.~J. Russell and P.~Norvig, ``Artificial intelligence: a modern approach
  (international edition),'' 2002.

\bibitem{fontes2015mininet}
R.~R. Fontes, S.~Afzal, S.~H. Brito, M.~A. Santos, and C.~E. Rothenberg,
  ``Mininet-wifi: Emulating software-defined wireless networks,'' in
  \emph{Network and Service Management (CNSM), 2015 11th International
  Conference on}.\hskip 1em plus 0.5em minus 0.4em\relax IEEE, 2015, pp.
  384--389.

\end{thebibliography}





\end{document}